\documentclass[aps,twocolumn,preprintnumbers,nofootinbib,superscriptaddress]{revtex4-1}
\usepackage{eurosym}
\usepackage{amsmath}
\usepackage{graphicx}
\usepackage{amsfonts}
\usepackage{array}
\usepackage{amsthm}
\usepackage{bm}
\usepackage{palatino}
\usepackage{mathpazo}
\usepackage{supertabular}
\usepackage{subfig}
\usepackage[breaklinks]{hyperref}
\usepackage{graphicx}
\usepackage{caption}
\usepackage[font=small,labelfont=bf,justification=raggedright]{caption}
\usepackage{color}
\usepackage{booktabs}
\usepackage{multirow}
\newcommand{\be}{\begin{equation}}
\newcommand{\ee}{\end{equation}}
\newcommand{\ba}{\begin{eqnarray}}
\newcommand{\ea}{\end{eqnarray}}
\newcommand{\bal}{\begin{align}}
\newcommand{\eal}{\end{align}}
\newcommand{\lb}{\label}

\newcommand{\bw}{\begin{widetext}}
\newcommand{\ew}{\end{widetext}}

\usepackage[utf8]{inputenc}
\usepackage{amssymb}
\usepackage{amsmath}
\usepackage{amsfonts}
\usepackage{graphicx, float}
\usepackage{graphicx, epsfig}
\usepackage{color}
\usepackage{enumerate}
\newcommand{\beq}{\begin{equation}}
\newcommand{\eeq}{\end{equation}}
\newcommand{\bea}{\begin{eqnarray}}
\newcommand{\eea}{\end{eqnarray}}

\begin{document}

\title{Constraining the Generalized Uncertainty Principle Through Black Hole Shadow  and Quasiperiodic Oscillations}
\author{Kimet Jusufi}
\affiliation{Physics Department, State University of Tetovo, Ilinden Street nn, 1200,
Tetovo, North Macedonia}
\affiliation{Institute of Physics, Faculty of Natural Sciences and Mathematics, Ss. Cyril
and Methodius University, Arhimedova 3, 1000 Skopje, North Macedonia}
\author{Mustapha Azreg-A\"{i}nou}
\affiliation{Ba\c{s}kent University, Engineering Faculty, Ba\u{g}l{\i}ca Campus, 06790-Ankara, Turkey}
	\author{Mubasher Jamil}
\email{mjamil@zjut.edu.cn (corresponding author)}
\affiliation{Institute for Theoretical Physics and Cosmology
	Zhejiang University of Technology
	Hangzhou, 310023 China}
\affiliation{Department of Mathematics, School of Natural Sciences (SNS),
National University of Sciences and Technology (NUST), H-12, Islamabad, Pakistan}

\author{Tao Zhu}
\affiliation{Institute for Theoretical Physics and Cosmology
	Zhejiang University of Technology
	Hangzhou, 310023 China}

\begin{abstract}
In this paper we study the effect of the Generalized Uncertainty Principle (GUP) on the shadow of GUP-modified Kerr black hole and the correspondence between the shadow radius and the real part of the quasinormal modes  (QNMs). We find that the shadow curvature radius of the GUP-modfied Kerr black hole is bigger compared to the Kerr vacuum solution and increases linearly monotonically with the increase of the GUP parameter. We then investigate the characteristic points of intrinsic curvature of the shadow from a topological point of view to calculate the the angular size for these curvature
radii of the shadow. To this end, we have used the EHT data for the M87* black hole to constrain the upper limits of the GUP parameter red and our result is $\beta<10^{95}$. Finally, we have explored the connection between the shadow radius and the scalar/electromagnetic/gravitational QNMs. The GUP-modified Kerr black hole is also used to provide perfect curve fitting of the particle oscillation upper and lower frequencies to the observed frequencies for three microquasars and to restrict the values of the correction parameter in the metric of the modified black hole to very reasonable bound $\beta<10^{77}$.
\end{abstract}
\maketitle

\section{Introduction}

Numerous geometrical and mathematical investigations concerning the interior regions of black hole suggest that not only general relativity but also quantum mechanics (more generally, known laws of physics) break down at the singularity. However, an amalgamation of general relativity and quantum mechanics or quantum gravity can predict novel features of black hole near the Planck length scale. Since the Heisenberg uncertainty principle (HUP) is not valid in its apparent form in the strong gravity, this law of physics gets gravitational corrections to investigate physics near high energy or short distance scales \cite{mage}.
In literatures, there are several versions of GUP from the HUP which are based either on some models of quantum gravity such as string theory, loop quantum gravity or phenomenology \cite{vage}. The well-known HUP obeys the following rule between the position and momentum operators: $[x,p]=i \hbar$, (or the uncertainty in the position and  momentum variables satisfies the inequality $\Delta x\Delta p>\hbar$) while a generalization of HUP is proposed to include both linear and quadratic terms in momentum being motivated from the string theory, doubly special relativity and phase space considerations $[x_i,x_j]=0=[p_i,p_j],$ as the following \cite{vagee,fras, j1} $$[x_i,p_j]= i\hbar\Big[ \delta_{ij}-\beta\Big( \delta_{ij}p+\frac{p_ip_j}{p} \Big)+\beta^2(\delta_{ij}p^2+3p_ip_j)\Big],$$ where $\beta=\beta' l_p/\hbar$ is a constant and $l_p$ denotes the Planck length. Therefore, when $i=j$, the last expression reduces to $[x,p]=i\hbar(1-2\beta p+4\beta^2 p^2).$ Similarly, the inequality representing the uncertainty in the position and momentum variables in the GUP is given by $\Delta x\Delta p>\hbar+\beta l_p^2 (\Delta p)^2$. In order to relate GUP with the black hole physics, one replaces $\Delta x$ with $r$ and $\Delta p$ with mass $M$ of the black hole. This consideration naturally yields $r>r'=\frac{\hbar}{M}+\beta l_p^2 M$.

A prediction using GUP considerations is that it can prevent the black hole evaporation completely near the Planck scale \cite{chen}. The backreaction effects involving the interaction of photons with electrons might stop the process of black hole evaporation as soon as the Planck mass is reached. The model proposes that the quantum gravity effects lead to the formation of black remnant \cite{wu,spa}. If the remnants exist in significant number in the universe than they can be a candidate of dark matter as well. The final remnant has a structure with a degenerate, extremal, horizon of radius of the order of the minimal length \cite{min}. The Hawking temperature of the black hole horizons gets GUP corrections which are meaningful only when black hole mass is greater than Planck mass otherwise the temperature gets imaginary. Thus the maximum possible black hole temperature is associated with the Planck scale.

An implication of GUP is that the ADM mass of black hole is modified as $\mathcal{M}=M+\beta M_{p}^2/M^2$, where $M$ is the bare mass of the black hole, $M_p$ is the Planck mass while $\beta>0$ is a constant \cite{Carr:2020hiz}. The authors of Ref.~\cite{Carr:2020hiz} proposed and analyzed the GUP corrected Schwarzscild, Reissner-Nordstrom and Kerr spacetimes. They showed that for a given value of $\beta$, there exists a critical charge and a critical spin above which the solutions bifurcate into sub-Planckian and super-Planckian phases, separated by a mass gap in which no black holes can form.  One the other hand, the strong gravity regime near a black hole is also thought to be a region that may help us to explore the quantum nature of the spacetime. This has stimulated a lot of attention on the phenomenological aspects of the GUP corrected black holes. For example, in literature, an upper bound on the GUP parameter is obtained using the data of the shadow of M87* central black hole \cite{m87}. By relating the GUP parameter with the deviation from the circularity of M87* black hole shadow, it is shown that $\beta<10^{90}$. To achieve this numerical value, the author employed a model of Kerr de Sitter black hole with a regular interior.  Besides, there are other set of constraints on the GUP parameter $\beta$ from the stellar dynamics and solar system tests such as the perihelion precession ($\beta<10^{69}$), pulsar periastron shift ($\beta<10^{71}$), deflection of light ($\beta<10^{78}$), gravitational waves ($\beta<10^{60}$), gravitational redshift ($\beta<10^{73}$), gravitational time delay ($\beta<10^{78}$) and geodetic precession ($\beta<10^{72}$) \cite{const}.  Our objective is to constrain the parameter $\beta$ appearing in the GUP corrected Kerr metric by making comparisons with the black hole shadow and quasi-periodic oscillations (QPOs).

In the strong gravity regime, the observational aspects of black holes are closely related to a narrow region not far from its event horizon, range from the photon sphere to the accretion region around the black hole. This region provides a great place to test the possible quantum gravity effects by using the electromagnetic observations including the black hole shadow and QPOs \cite{bambi, bambi2}.  A black hole shadow is a two-dimensional dark zone in the celestial sphere caused by the strong gravity of the black hole. The shape and size of the shadow mainly depend on the geometry of the black hole spacetime \cite{eht}. With this reason, by observing both the shape and the size of the shadow, one is able to extract valuable information, including the GUP effects in the black hole spacetime. For QPOs, it is a phenomenon arising in the X-ray radiation from black holes or compact objects which are accreting material from a stellar companion. These X-ray radiation is emitted from the innermost regions of the accretion region thus can provide another powerful way to explore the strong gravity regime of the black hole spacetime and test non-Kerr spacetime \cite{bambi, yuan, bambi2}. It is expected that the GUP can provide important effects in the QPOs so one can use the QPOs observations to constrain the GUP parameter. With these motivations, in this paper, we study the effects of the GUP in both the shadow and QPOs by considering the GUP-corrected Kerr black hole. With the recent observations, we also derive the observational constraints on the GUP parameter.

The paper is structured as follows:  In Sec. II, after a brief review of GUP-modified Kerr black hole, we investigate its shadow and calculate the curvature radius of its boundary. In Sec. III, we determine the constraints on the GUP parameter using the astronomical data related with the shadow of M87* supermassive black hole. In Sec. IV, we investigate the connection between quasinormal modes and the shadow radius. In Sec. V, we study the quasi-periodic oscillations (QPOs) in order to constrain GUP parameter as another method. Finally in Sec. VI, we comment on our results in the conclusion. We choose the units $c = \hbar = G = 1$ throughout the manuscript except Sec. V and occasionally set Planck mass $M_p=1$. 

\section{GUP-modified Kerr black hole and its shadow}

Carr {\it et. al.}, explored the question what kind of short or large distance corrections to the radial length parameter $r$ should be considered as one goes from a given large distance to a short distance scale and vice versa  \cite{Carr:2020hiz}. Using the two fundamental length scales namely the short distance Compton wavelength $r_{\rm C}$ and the comparatively long distance Schwarzschild radius $r_{\rm S}$, for relating quantum and classical black hole physics, they proposed that as one goes from Schwarzschild radius to the Compton length scale, the corresponding correction to $r_{\rm C}$ should include of order $\alpha M$ and similarly if one goes from the Compton wavelength scale to the Schwarzschild scale than the necessary correction should be of order $\beta /M$. To relate and amalgamate the two modified length scales, they suggested that the most appropriate length scale should be $r_{\rm CS}=\frac{\hbar\beta}{Mc}+\frac{2GM}{c^2}$. They further proposed that the bare mass of the Schwarzschild black hole, consequently gets short distance corrections, as given below in Eq. (3). We consider below a Kerr black hole with short distance correction and explore its effects on the shadow of black hole. 

We can proceed with the line element of the Kerr spacetime given by
\begin{eqnarray}\notag
ds^2 & =& -\left(1-\frac{r_{\rm S} r}{\rho^2}\right) dt^2 + \frac{\rho^2}{\Delta} dr^2-
\frac{2 r_{\rm S} r \mathrm{a} \sin^2\theta }{\rho^2}  ~dt~d\phi\\
&&+   \rho^2~ d\theta^2 +\left(r^2 + a^2  + \frac{r_{\rm S}\, r\, a^2}{\rho^2} \sin^2\theta\right) \sin^2\theta~ d\phi^2,\label{metric}
\end{eqnarray}
where $M$ and $a$ denote the bare mass and spin of the black hole while other metric parameters are defined as
\begin{equation}
r_{\rm S} = 2M,~~a = \frac{J}{M},~~\rho^2 = r^2 + a^2 \cos^2\theta,~~\Delta = r^2 - r_{\rm S} r + a^2 \, .
\end{equation}
In order to write the GUP-corrected Kerr metric, we need to replace the black hole bare mass $M$ by the ADM mass $\mathcal M$ defined by \cite{Carr:2020hiz}:
\begin{equation}
M\to \mathcal{M} \equiv M+\frac{\beta}{2 M}.
\end{equation}
In that case, we have the following relations
\begin{eqnarray}
r_{\rm S} &=&2M \zeta^{-1},
\\ a\to \mathrm{a} &=&a \zeta,~~~ \mathcal M=M\zeta^{-1},\\
\Delta &=&r^2-2M \zeta^{-1} r+a^2 \zeta^2,
\end{eqnarray}
where we have defined
\begin{eqnarray}
\zeta\equiv\left(1+\frac{\beta}{2 M^2}\right)^{-1}.
\end{eqnarray}

In order to find the contour of a black hole shadow,  we need to separate the null geodesic equations in the general rotating spacetime metric (1) using the Hamilton-Jacobi equation given by
\begin{equation}
\frac{\partial \mathcal{S}}{\partial \sigma}=-\frac{1}{2}g^{\mu\nu}\frac{\partial \mathcal{S}}{\partial x^\mu}\frac{\partial \mathcal{S}}{\partial x^\nu},
\label{eq:HJE}
\end{equation}
where $\sigma$ is the affine parameter, $\mathcal{S}$ is the Jacobi action. For this purpose we can express the action in terms of known constants of the motion as follows
\begin{equation}
\mathcal{S}=\frac{1}{2}m ^2 \sigma - E t +l \phi + \mathcal{S}_{r}(r)+\mathcal{S}_{\theta}(\theta),
\label{eq:action_ansatz}
\end{equation}
where $m$ is the mass of the test particle, $E=-p_t$ is the conserved energy and $l=p_\phi$ is the conserved angular momentum. After substituting $m=0$ one can obtain the following equations of motion 
\begin{eqnarray}\notag
 \rho^2\frac{dt}{d\lambda}&=&a\zeta(l-a\zeta E\sin^{2}\theta)
       +\frac{r^{2}+a^{2}\zeta^2}{\Delta}\left[E\,(r^{2}+a^{2}\zeta^2)-a\zeta\,l\right],\\\notag
 \rho^2 \frac{dr}{d\lambda}&=& \pm \sqrt{\Re},\\\notag
 \rho^2 \frac{d\theta}{d\lambda}&=&\pm \sqrt{\Theta},\\
\rho^2\frac{d\phi}{d\lambda}
     &=&(l\csc^{2}\theta-a\zeta\,E)+\frac{a\zeta}{\Delta}\left[E(r^{2}+a^{2}\zeta^2)-a\zeta\,l\right],
\end{eqnarray}
where $\sigma$ is the affine parameter, $l$ is the angular momentum of the photon, $E$ is the energy of the photon and $\mathcal{K}$ is the Carter constant. In addition we have introduced
\begin{eqnarray}
 {\Re}&=&\left(a^2\zeta^2 \,E-a\zeta\,l+E\, r^2\right)^2-\Delta
   \left[\mathcal{K}+(l-a\zeta\, E)^2\right],\\
 \Theta&=&\mathcal{K}-(l\csc\theta-a\zeta\,
   E \sin\theta)^2+(l-a\zeta\,E)^2.
\end{eqnarray}

The size and shape the black hole shadow is determined by the unstable circular photon orbits satisfying the following conditions
\begin{equation}\lb{condition}
\Re(r)=0,\;\; \frac{d\Re(r)}{dr} =0 ,\;\;\; \frac{d^2 \Re(r)}{dr^2} >0.
\end{equation}
By using this condition the circular orbit radius $r_{ph}$ of the photon can be obtained and the parameters $\xi\equiv l/E$ and $\eta\equiv\mathcal{K}/E^{2}$ can thus be expressed as
where $X(r)=(r^2+a^2\zeta^2)$, and $\Delta(r)$ is defined by Eq. (3), while $\mathcal{K}$ is known as the Carter separation constant. From these conditions one can show that the motion of the photon can be determined by the following two impact parameters 
\begin{eqnarray}
\xi&=&\frac{X_{ph}\Delta'_{ph}-2\Delta_{ph}X'_{ph}}{a\zeta\Delta'_{ph}},\\\notag
\eta&=&\frac{4a^2\zeta^2 X'^2_{ph}\Delta_{ph}-\left[\left(X_{ph}-a^2\zeta^2\right)\Delta'_{ph}-2X'_{ph}\Delta_{ph} \right]^2}{a^2\zeta^2\Delta'^2_{ph}}.
\end{eqnarray}

One constraint for the value of the photon's circular orbit radius is ${\Re(r_{ph})>0}$.
The shape of the shadow seen by an observer at spatial infinity can be obtained from the geodesics of the photons and described by the celestial coordinates
\begin{eqnarray}\label{celkerrone}
x&=&-\xi\csc\theta_{0},\\
 y&=&\pm\sqrt{\eta+a^{2}\zeta^2 \cos^{2}\theta_{0}-\xi^{2}\cot^{2}\theta_{0}},
\end{eqnarray}
\begin{figure*}
\includegraphics[width=8.5cm]{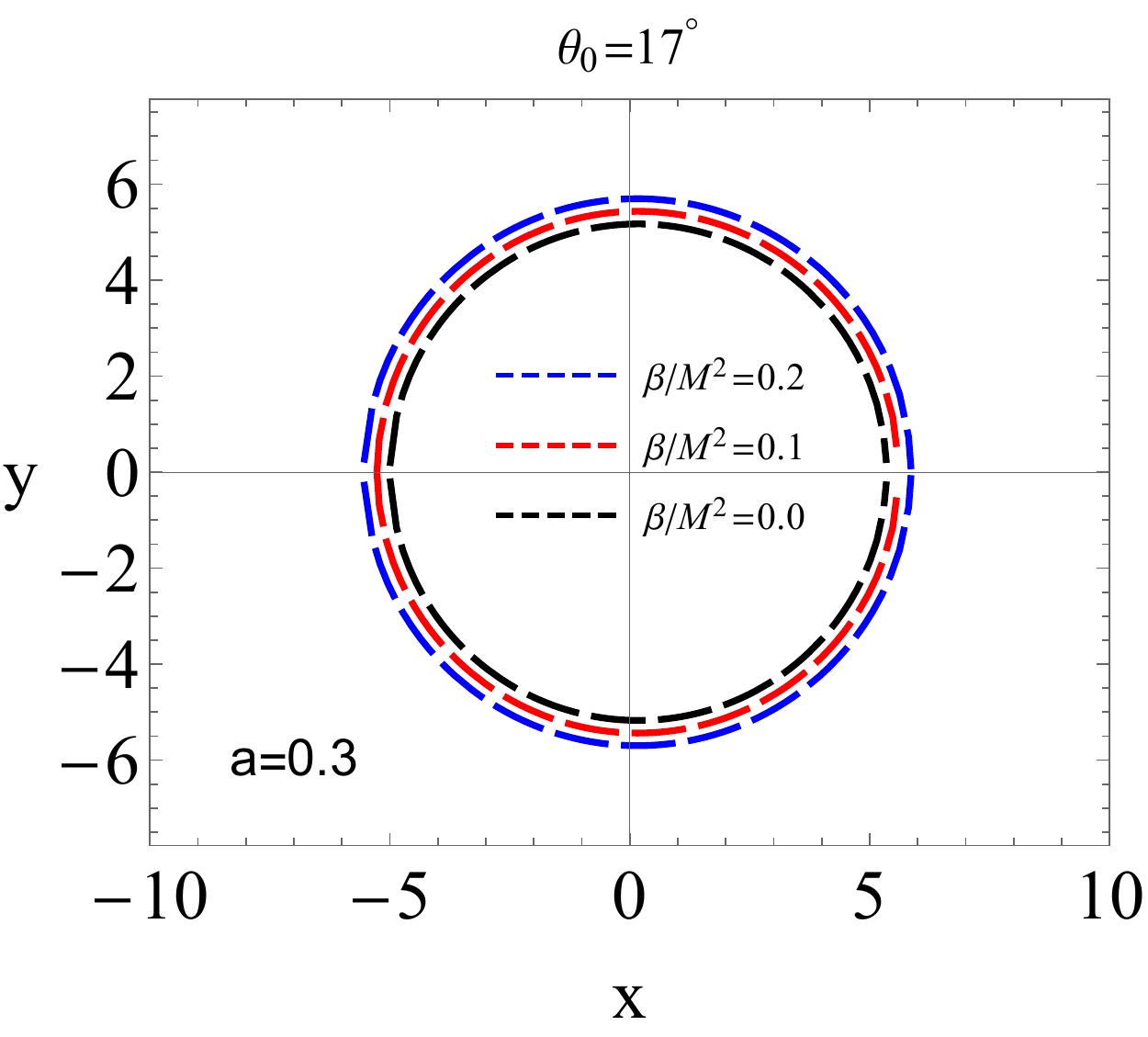}
\includegraphics[width=8.5cm]{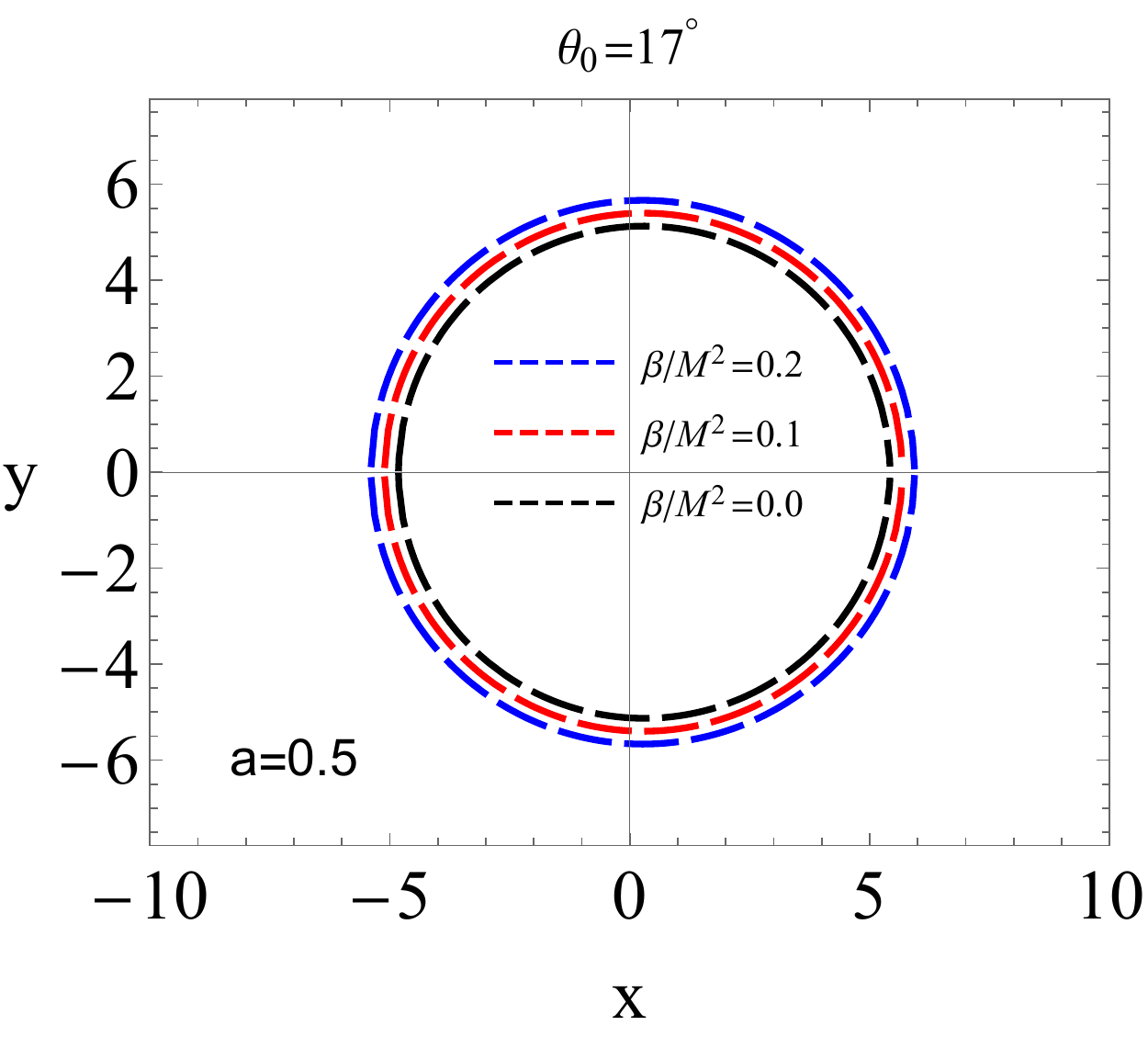}
\includegraphics[width=8.5cm]{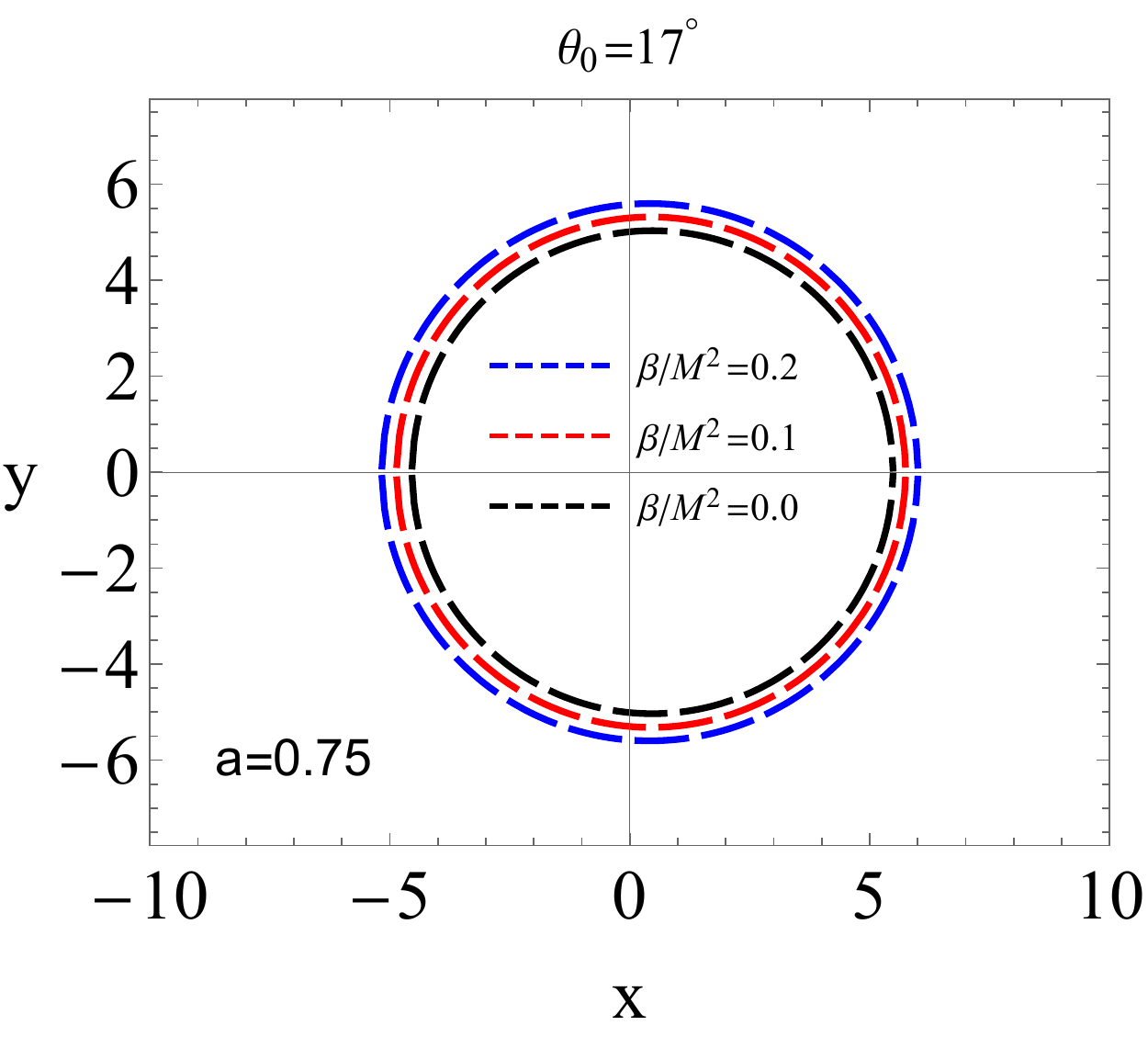}
\includegraphics[width=8.5cm]{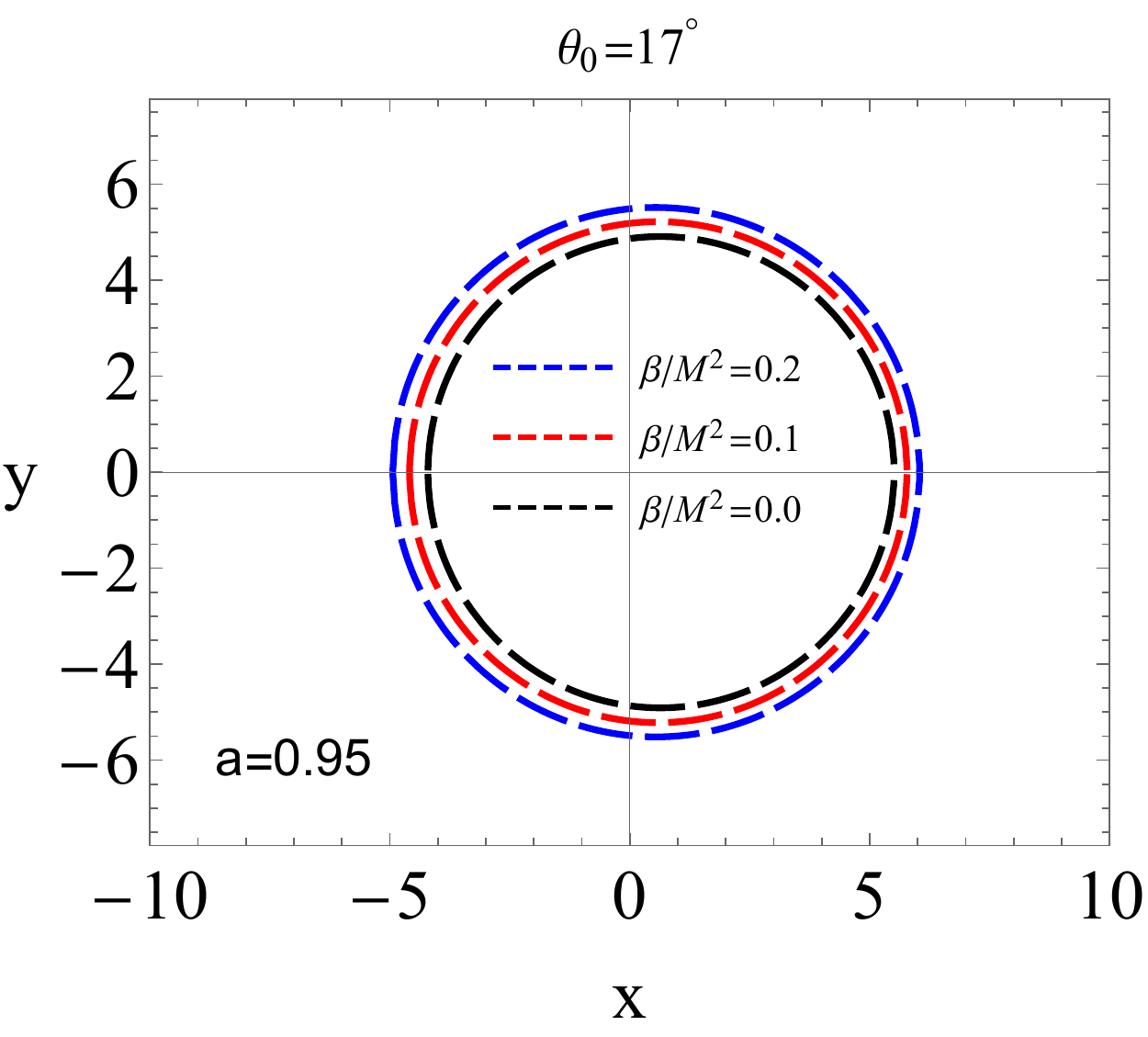}
\caption{The shape of shadow for the GUP-modified Kerr BH using the inclination angle $\theta_0=17^0$ for different values of $\beta$ and $a$. We have set $M=1$ in all cases. Note that $a$ has units of M, while $\beta$ has units of $M^2$. } \label{A}
\end{figure*}
In Fig. \ref{A} we plot the shape of the shadow by varying the GUP parameter. In our plots we are going to use an inclination angle   $\theta_0=17^o$ which is the value chosen by the EHT Collaboration to study  the shadow of M87 * central black hole \cite{eht}.  Furthermore, this angle has been obtained by studying the emission jet from the black hole (see \cite{jet}). It is observed from Fig. \ref{A} that the boundary of shadow does not deviate from the complete circle even when spin parameter is increased to an extreme value. However, significant deviation in the geometry of black hole shadow appears when the inclination angle is chosen to be more than $86^o$, which may be adopted for theoretical reasons \cite{bambi2}.  We observe that an increase of $\beta$, increases the shadow radius. Thus in general, for any $\beta\geq 0$, the shadow of GUP-modified black hole is bigger compared to the Kerr vacuum black hole. 

 Wei and collaborators studied a geometrical and a topological property of the Kerr black hole \cite{Wei:2018xks}. Assuming the black hole has a reflection symmetry about the y-axis, referred as the $Z_2$ symmetry in the two dimensional plane, and parameterising the boundary of the shadow by the curve $\{x(\lambda),y(\lambda) \}$, they calculated the total length of the boundary of shadow. The $x$ and $y$ parameters are also referred as the celestial coordinates used by an observer at spatial infinity. Given the curve, they calculated the radius of curvature of the curve. In general, the boundary of the black hole's shadow is a closed curve which may deviate from a circle by the variation of the parameters. Since a spinning black hole also contains a naked singularity in the special case, $a>|M|$, the shadow of a naked singularity constitutes an arc which shrinks in size if spin gets sufficiently large. The authors deduced the following expression of curvature radius for the shadow of a Kerr black hole  \cite{Wei:2018xks}
\begin{widetext}
\begin{eqnarray}
 R=\frac{64\mathcal{M}^{1/2}(r_{0}^{3}-\mathrm{a}^{2}r_{0}\cos^{2}\theta_{0})^{3/2}\left[r_{0}(r_{0}^{2}-3\mathcal{M}r_{0}+3\mathcal{M}^{2})-\mathrm{a}^{2}M^{2}\right]}
   {(r_{0}-\mathcal{M})^{3}\left[3(8r_{0}^{4}-\mathrm{a}^{4}-8\mathrm{a}^{2}r_{0}^{2})-4\mathrm{a}^{2}(6r_{0}^{2}+\mathrm{a}^{2})\cos(2\theta_{0})
    -\mathrm{a}^{4}\cos(4\theta_{0})\right]},
\label{Rad00}
\end{eqnarray}
where $r_0$ denotes the boundary of the photon sphere. For a Schwarzschild black hole, we have $r_0=3M$. If we use the scaling 
\begin{eqnarray}
 \mathcal{M} \to M\zeta^{-1},\,\,\,\mathrm{a} & \to & a \zeta ,
\end{eqnarray}
we obtain the following result for the local curvature radius 
\begin{eqnarray}
 R=\frac{64M^{1/2}\zeta^{-1/2}(r_{0}^{3}-a^{2}\zeta ^2r_{0}\cos^{2}\theta_{0})^{3/2}\left[r_{0}(r_{0}^{2}-3M\zeta^{-1}r_{0}+3M^{2}\zeta^{-2})-a^{2}M^{2}\right]}
   {(r_{0}-M\zeta^{-1})^{3}\left[3(8r_{0}^{4}-a^{4}\zeta ^4-8a^{2}\zeta ^2r_{0}^{2})-4a^{2}\zeta ^2(6r_{0}^{2}+a^{2}\zeta^2)\cos(2\theta_{0})
    -a^{4}\zeta ^4\cos(4\theta_{0})\right]},
\label{Rad00}
\end{eqnarray}
\end{widetext}

\begin{figure}
\includegraphics[width=7cm]{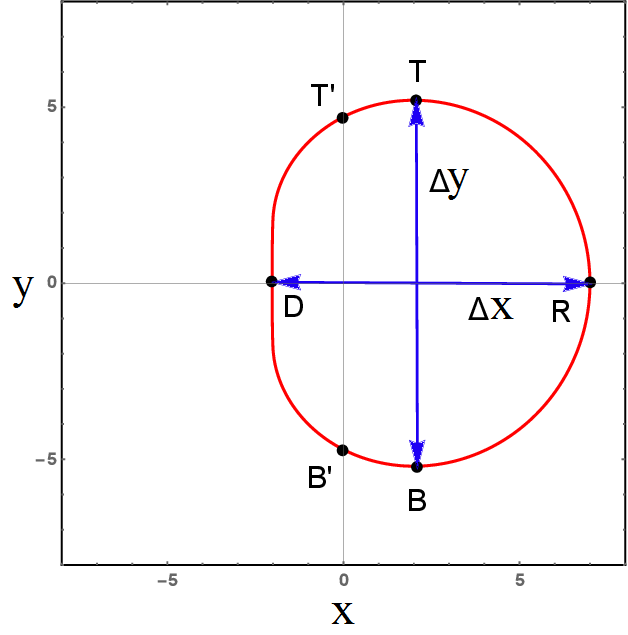}
\caption{The shape of shadow and the characteristic points of intrinsic curvature of the shadow (picture adopted from the Ref. \cite{Wei:2018xks}).  Notice the symmetry of the closed curve about the y-axis, commonly referred as $Z_2$ symmetry. } \label{BB}
\end{figure}

\begin{figure*}
\includegraphics[width=8.5cm]{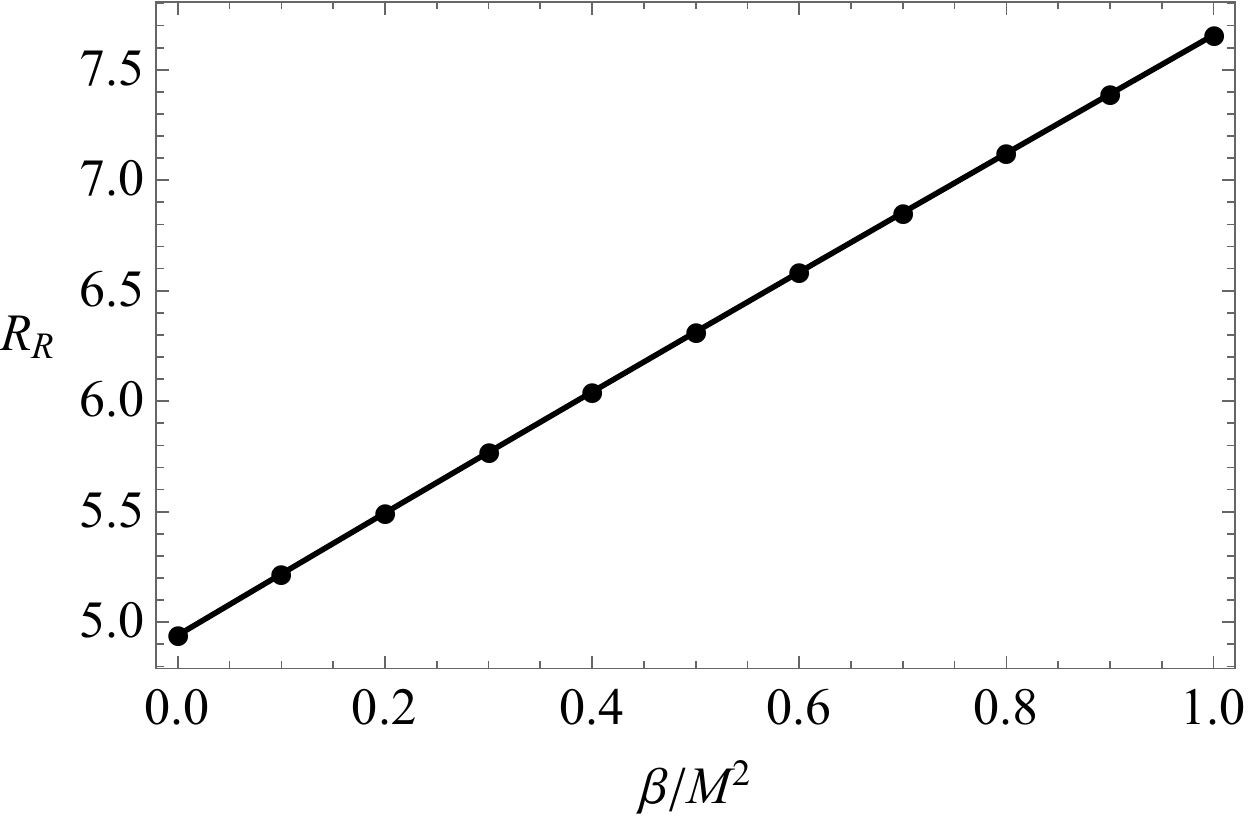}
\includegraphics[width=9.3 cm]{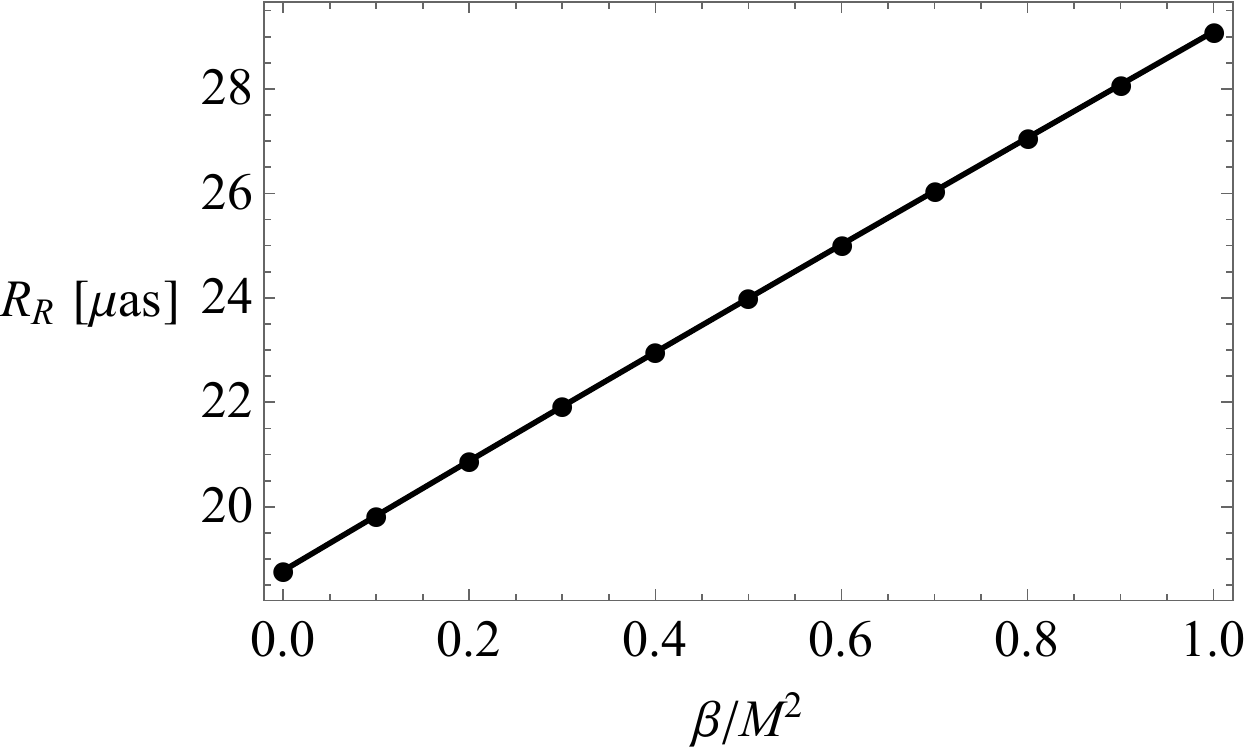}
\includegraphics[width=8.5cm]{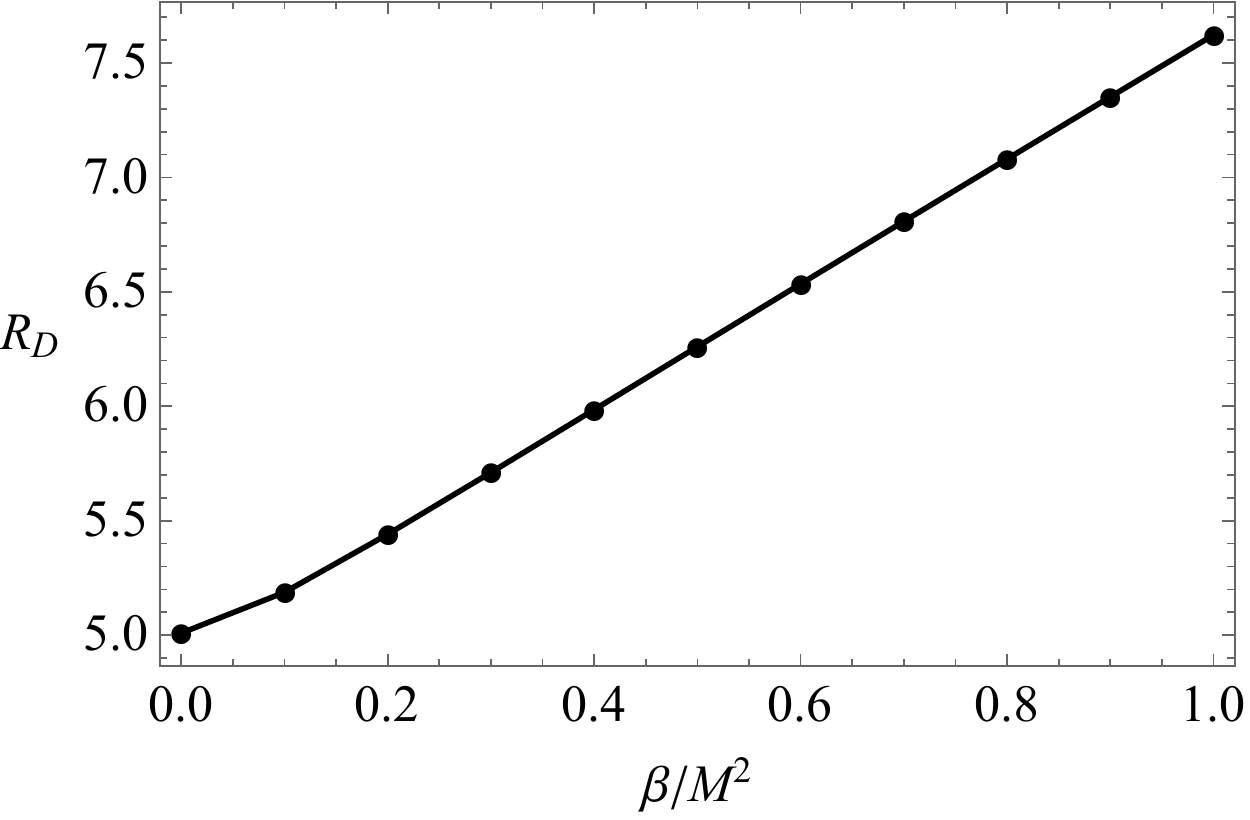}
\includegraphics[width=9.3 cm]{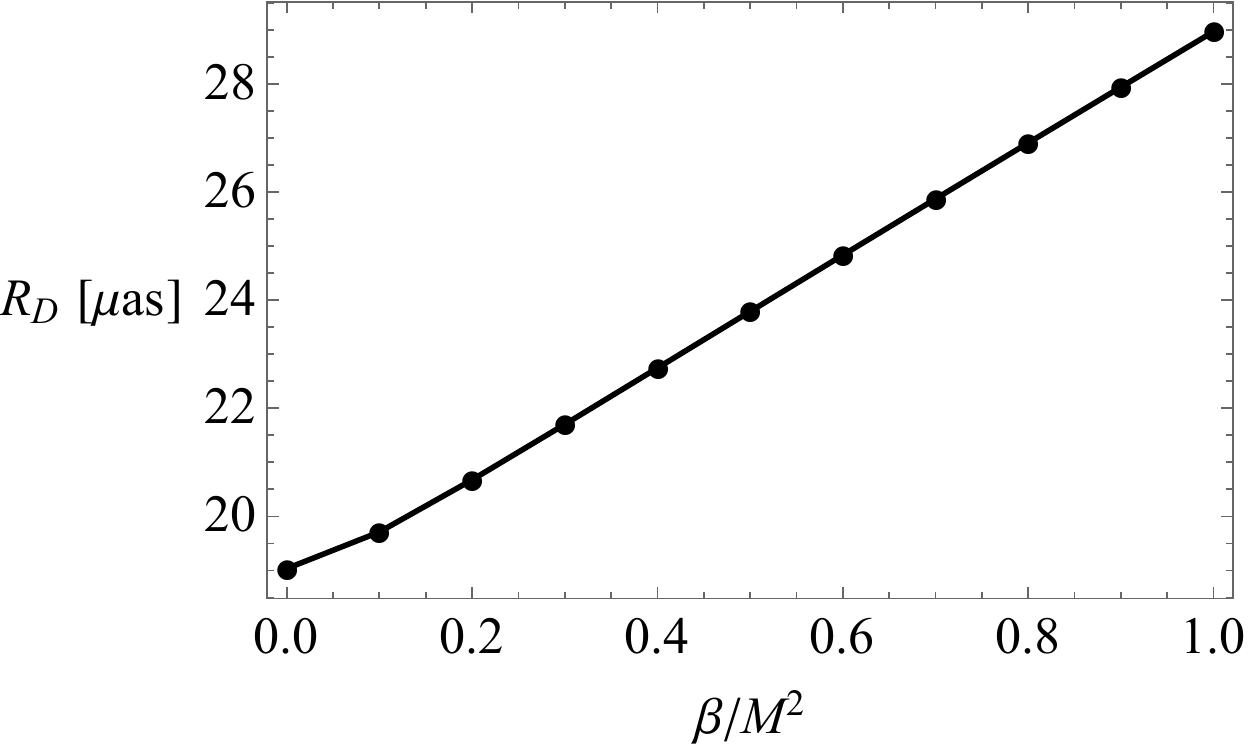}
\includegraphics[width=8.5cm]{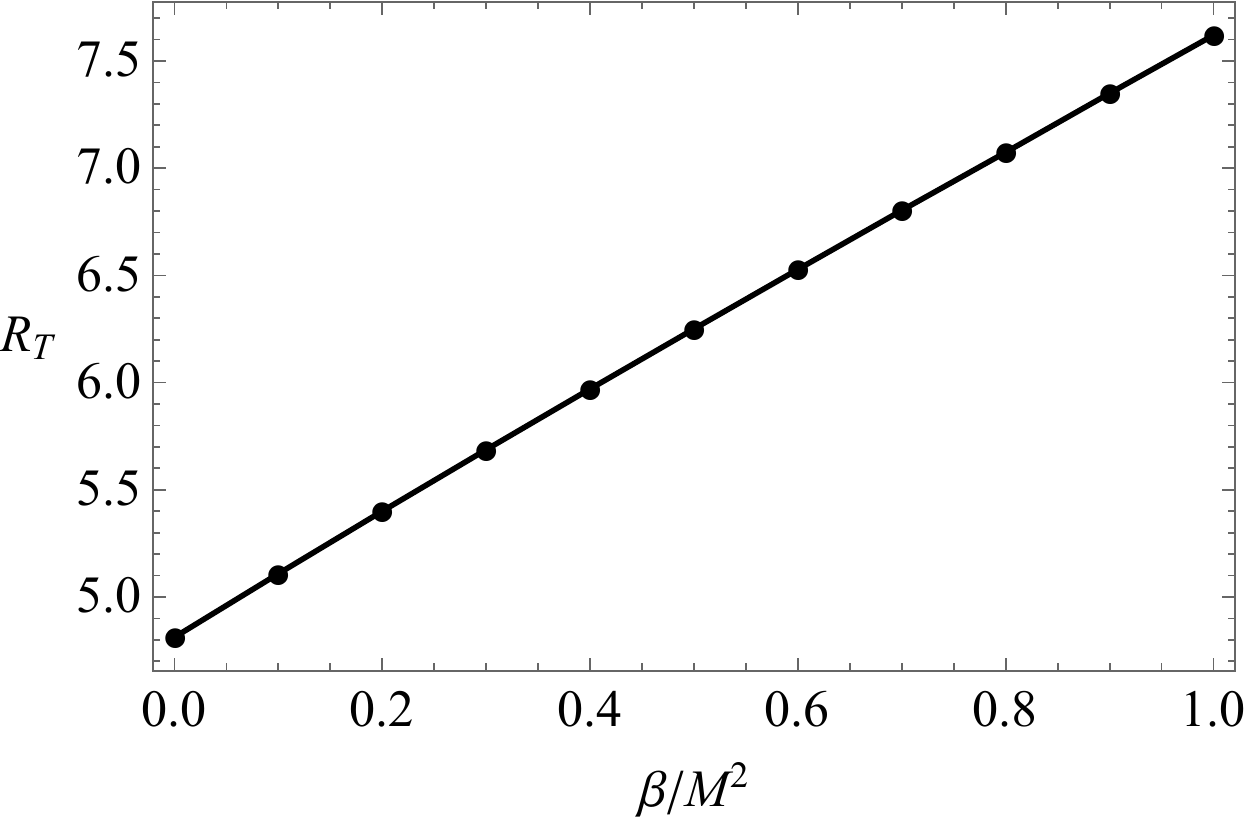}
\includegraphics[width=9.3 cm]{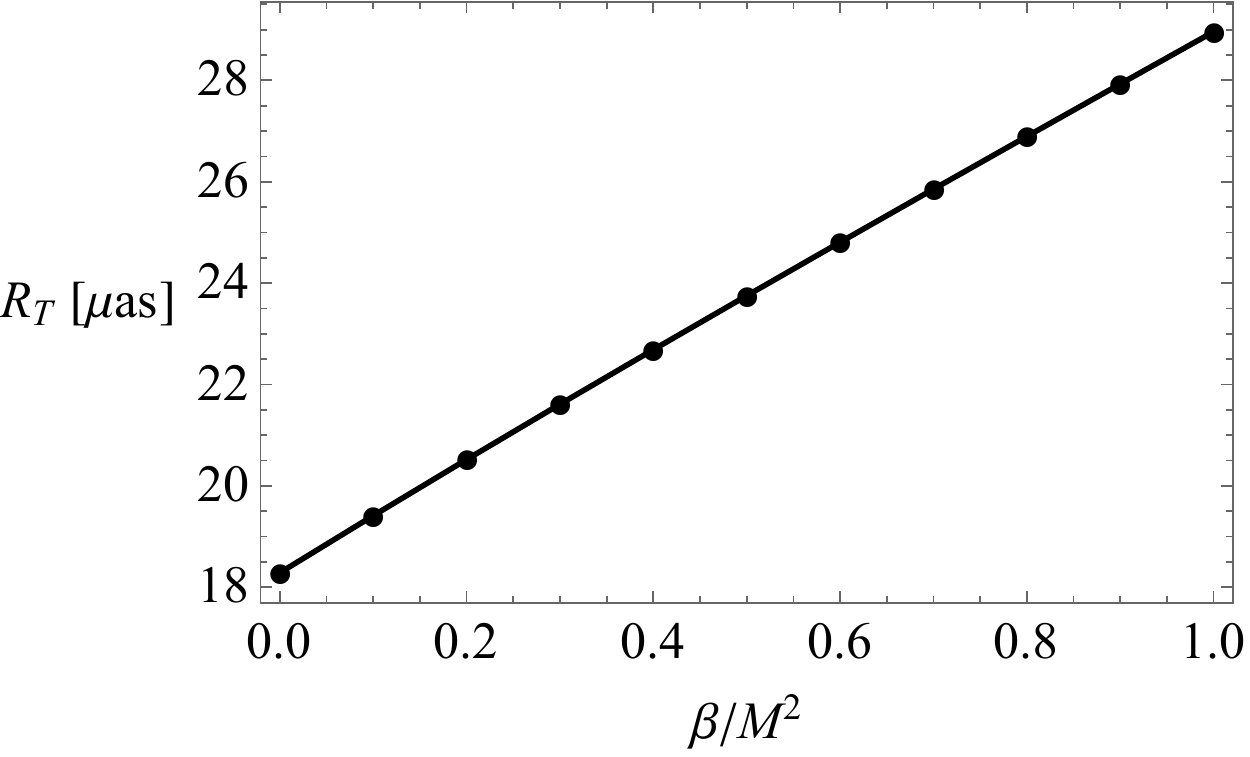}
\caption{The plots the local curvatures of the shadow asa function of $\beta$ for the three points: R, D, and T. We have set $M=1$, $a/M=0.95$ and $\theta_0=17^0$.  } \label{B1}
\end{figure*}

\begin{figure*}
\includegraphics[width=8.7 cm]{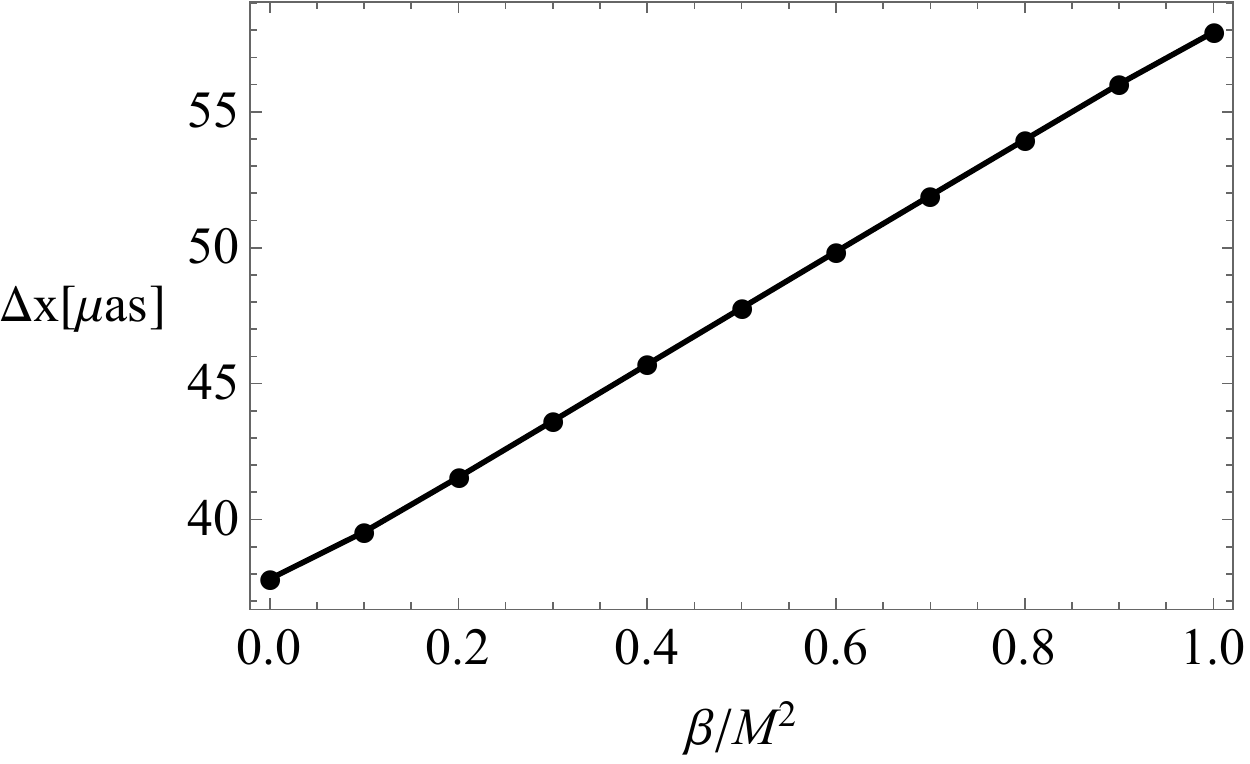}
\includegraphics[width=8.7 cm]{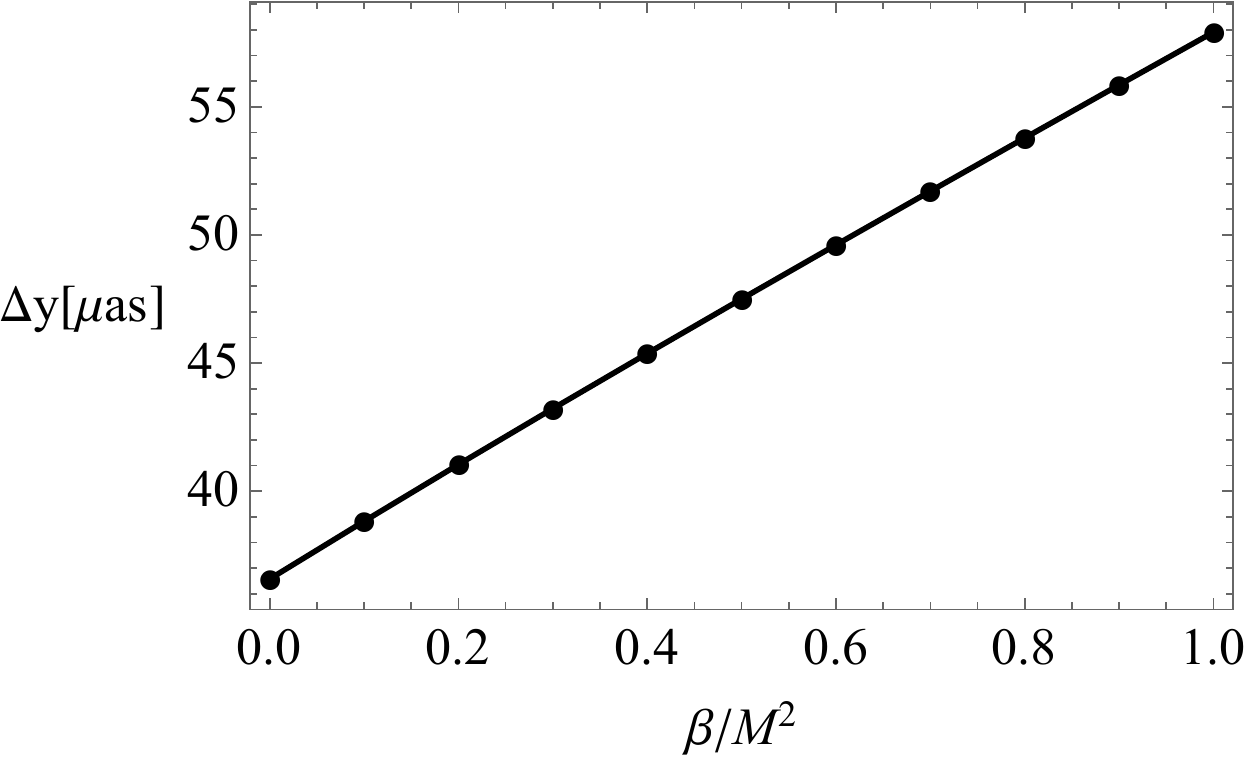}
\caption{The shape of shadow for the GUP corrected Kerr BH using the inclination angle $\theta_0=17^0$ for different values of $\beta$ and $a$. We have set $M=1$.  } \label{B2}
\end{figure*}

In what follows we are going to use the last equation to evaluate the  intrinsic curvature of the shadow in three characteristic points: D, R and T, respectively, see Fig. \ref{BB}.  In particular we are going to compute the the horizontal and vertical angular size for these curvature radii of the shadow, noted as $\Delta x$ and $\Delta y$, respectively. In Fig. \ref{B1} and Fig. \ref{B2}, we display the plots for the local curvature and the angular size for these curvature radii of the shadow as a function of $\beta$ by fixing $\theta_0$ and $a$. We see that the local curvature monotonically increases with the increase of $\beta$.
\begin{figure*}
\includegraphics[width=8.5cm]{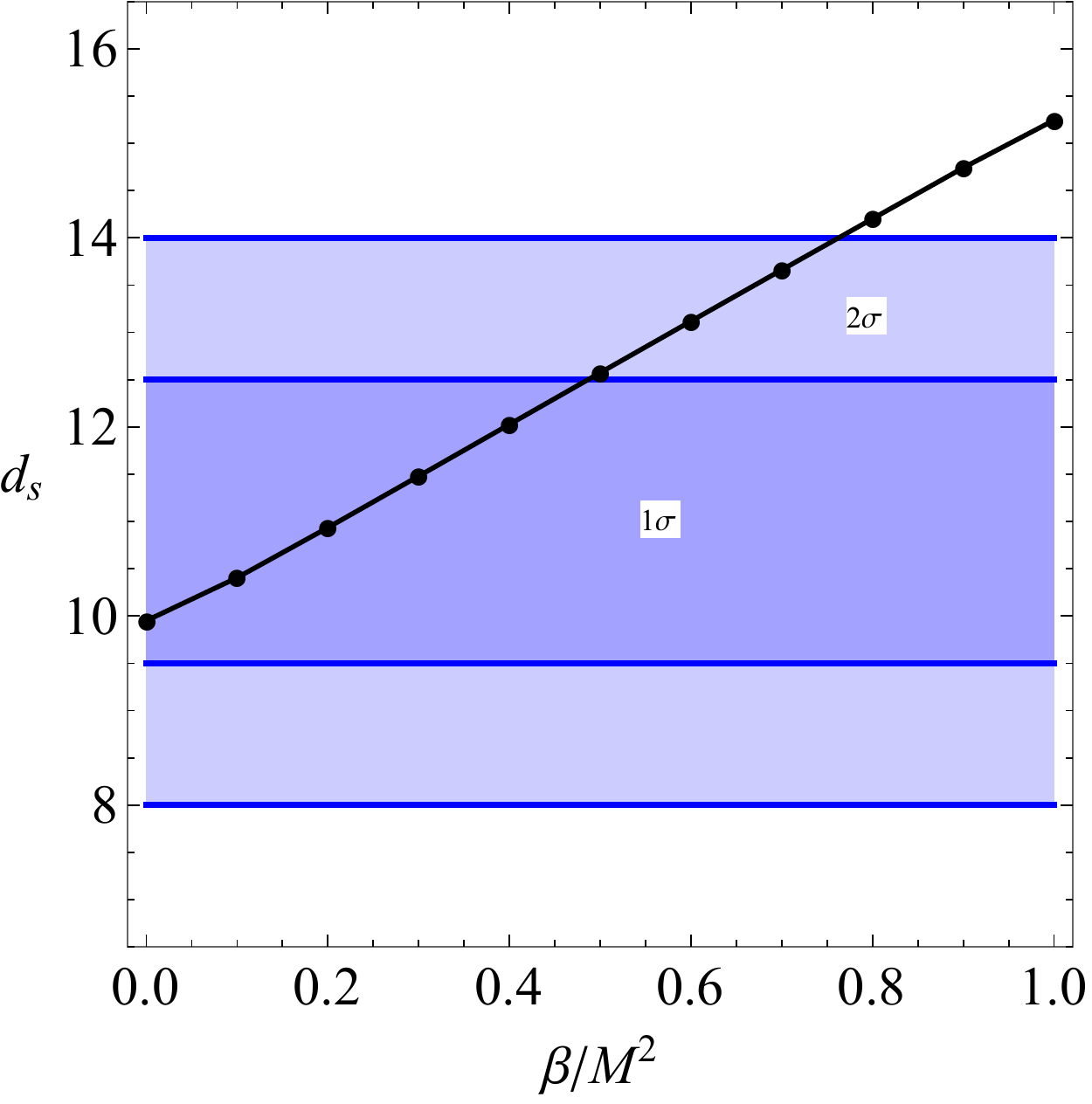}
\includegraphics[width=8.5cm]{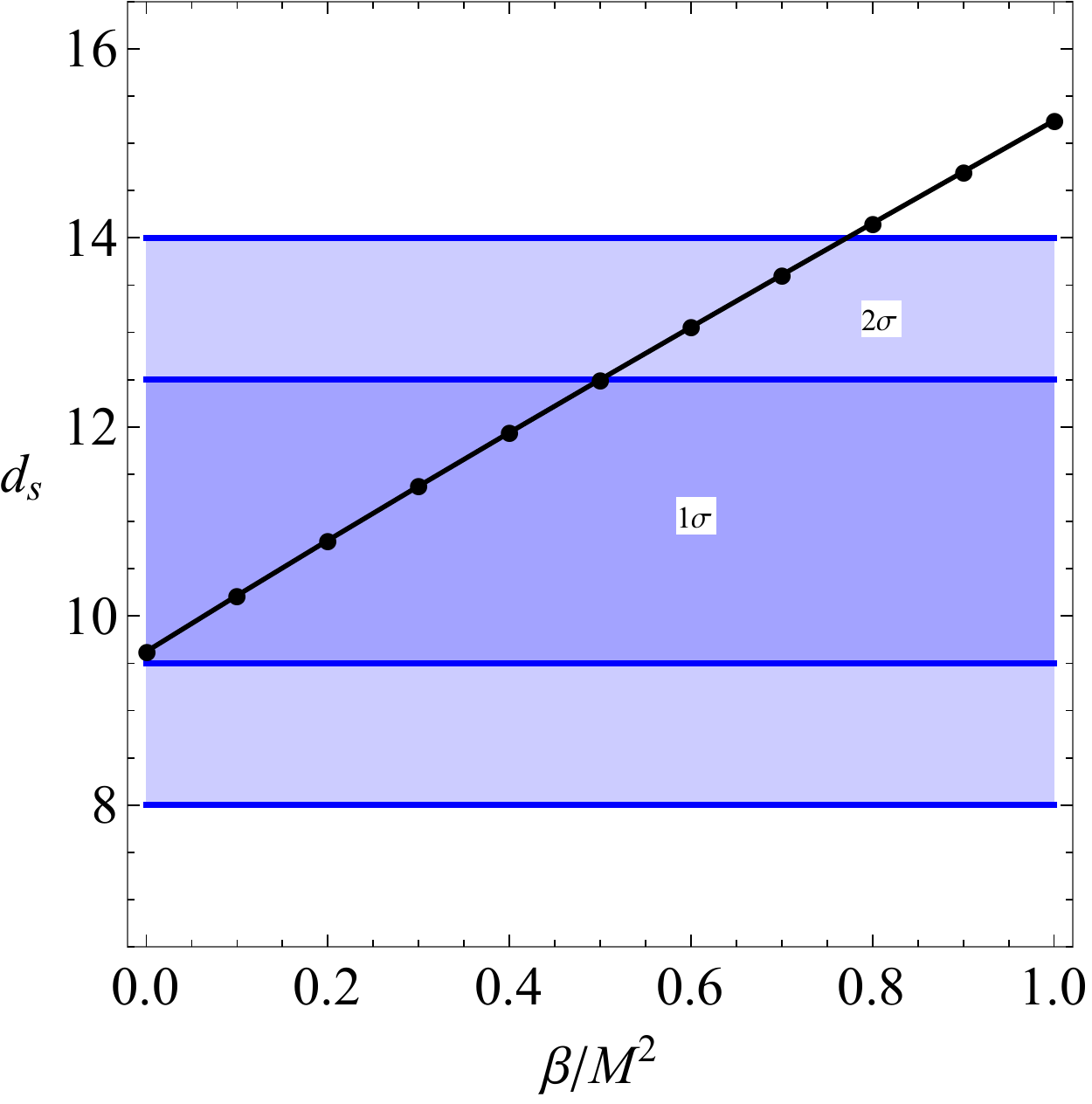}
\caption{Left panel: The regions of parameter space of the diameter of the shadow and the GUP parameter within $1\sigma $ and $2\sigma$ uncertainties, respectively.  Right panel: The regions of parameter space and the GUP parameter for the second case. We have set the inclination angle $\theta_0=17^0$ along with $a/M=0.95$ measured in units of the M87* black hole $M$. } \label{B3}
\end{figure*}

\section{Observational constraints}

 We can use the reported angular size of the shadow in the M87* center detected by the EHT  $\theta_s = (42 \pm 3)\mu as$, along with the distance to M87* given by  $D = 16.8^{+0.8}_{-0.7} $ Mpc, and the
mass of M87* central object $M = (6.5\pm 0.9) \times 10^9$ M\textsubscript{\(\odot\)} to constrain the GUP parameter. To do so, we are  going to consider two cases:\\
$a)$ In the first case, we identify the diameter of shadow in units of mass $d_{M87}$  with the horizontal angular size $\Delta x$. \\
$b)$ In the second case, we identify  the diameter of shadow in units of mass $d_{M87}$  with the vertical angular size $\Delta y$. \\ 

Now we can use \cite{Allahyari:2019jqz}
\begin{eqnarray}
d_{M87}=\frac{D \,\theta_s}{M}=11.0 \pm 1.5.
\end{eqnarray}

Within $1\sigma $ confidence we have the interval $9.5 \lesssim d_{M87} \lesssim 12.5$, whereas within $2\sigma $ uncertainties we have $8 \lesssim d_{M87} \lesssim14 $ \cite{Allahyari:2019jqz}. In Fig. 5 we show the regions of parameter space of the diameter of the shadow and the GUP parameter $\beta$ for two cases. In the first case, within $1\sigma$ confidence, we find an upper limit of the GUP parameter $\beta/M^2 \lesssim 0.77$. On the other hand, within $2\sigma$ confidence, we find an upper limit $\beta/M^2 \lesssim 0.49$
In the second case, we identify  the diameter of the shadow in units of mass $d_{M87}$  with the vertical angular size $\Delta y$. We find almost similar results, namely within $1\sigma$ confidence, we find the upper limit to be $\beta/M^2 \lesssim 0.5 $. On the other hand, within $2\sigma$ confidence, we find the interval $\beta/M^2 \lesssim 0.78 $. In other words, the results are the same. Notice also that case of negative $\beta$ is probably unphysical, therefore  we rule out negative values of $\beta$.  This means that within $2\sigma $,  we have the upper and lower limit of the GUP parameter in terms of the interval $0 \lesssim \beta/M^2 \lesssim 0.78 $. The GUP parameter has been estimated to be $\beta/M^2<0.78$, but measured in units of black hole mass squared. However, in order $\beta $ to be dimensionless, we have to restore the Planck mass $[M_P=2.2 \times 10^{-5}$ g].  That means
$\beta \lesssim 0.78 \times M^2/M_P^2$. Given the fact that the M87 black hole mass is $6.5 \times  10^9 M$\textsubscript{\(\odot\)}, we obtain a upper bound $\beta \lesssim 2.7 \times 10^{95}$. Since this quantity scales with the black hole mass, we see that as the black hole mass decreases, this bound should decrease as well. That's why quantum systems are better in constraining GUP. For primordial black holes we can take the mass about $10^{15}$ g and obtain $\beta \lesssim 1.6 \times 10^{39}$. According to GUP, there should be some final size or so called remnant mass where the Hawking evaporation stops in such case the black hole mass is of the order of Planck mass, hence  $\beta \lesssim 0.78$.

\section{Connection between shadow radius and QNMs}
\label{qnms}
Now we proceed further to investigate the correspondence between the 
radius of the black hole shadow and the real part of the QNMs frequency. 
It has been already known that the real part of the 
the QNMs frequencies is related to the angular velocity of the unstable 
null geodesic in the eikonal limit 
\cite{cardoso}. Moreover, the imaginary part of the QNMs frequencies is 
related to the Lyapunov exponent that determines the instability time 
scale of the orbits. This can be easily understood by the following 
equation \cite{cardoso} 
\begin{equation}
	\omega_{QNM}=\Omega_c l -i \left(n+\frac{1}{2}\right)|\lambda|, 
	\label{41}
\end{equation}
where $\Omega_c$ is the angular velocity at the unstable null 
geodesic, and $\lambda$ denotes the Lyapunov exponent. Furthermore, 
this correspondence is expected to be valid not only for the static 
spacetimes but also for the stationary ones. On the other hand, 
Stefanov {\it et al.} \cite{Stefanov:2010xz} showed a connection 
between the QNMs frequencies and the strong gravitational lensing of 
the spherically symmetric black holes spacetime. Most recently, one 
of the authors of this paper pointed out that the following relation 
relates the real part of the QNMs frequencies and the shadow radius 
(see for details \cite{Jusufi:2019ltj,Liu:2020ola})
\begin{equation}
	\omega_{\Re} = \lim_{l \gg 1} \frac{l}{R_s}, \label{k1}
\end{equation}
which is precise only in the eikonal limit having large values of 
multipole number $l$. Here $R_s$ denotes the radius of the black 
hole shadow. Hence, we can quickly rewrite the expression \eqref{41} 
as 
\begin{equation}
	\omega_{QNM}=\lim_{l \gg 1} \frac{l}{R_s} 
	-i \left(n+\frac{1}{2}\right)|\lambda|.
\end{equation}
\begin{figure*}
\includegraphics[width=8.5cm]{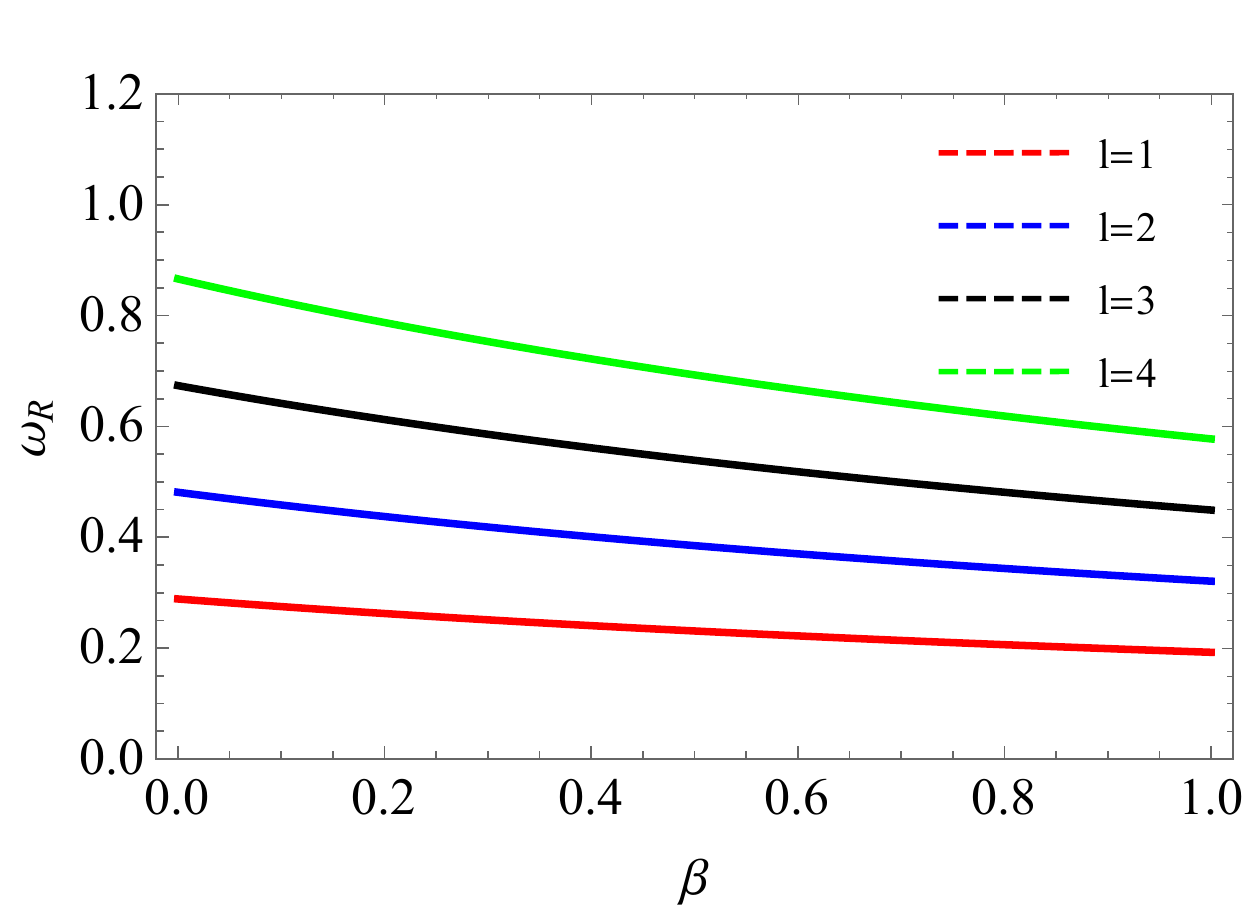}
\includegraphics[width=8.5cm]{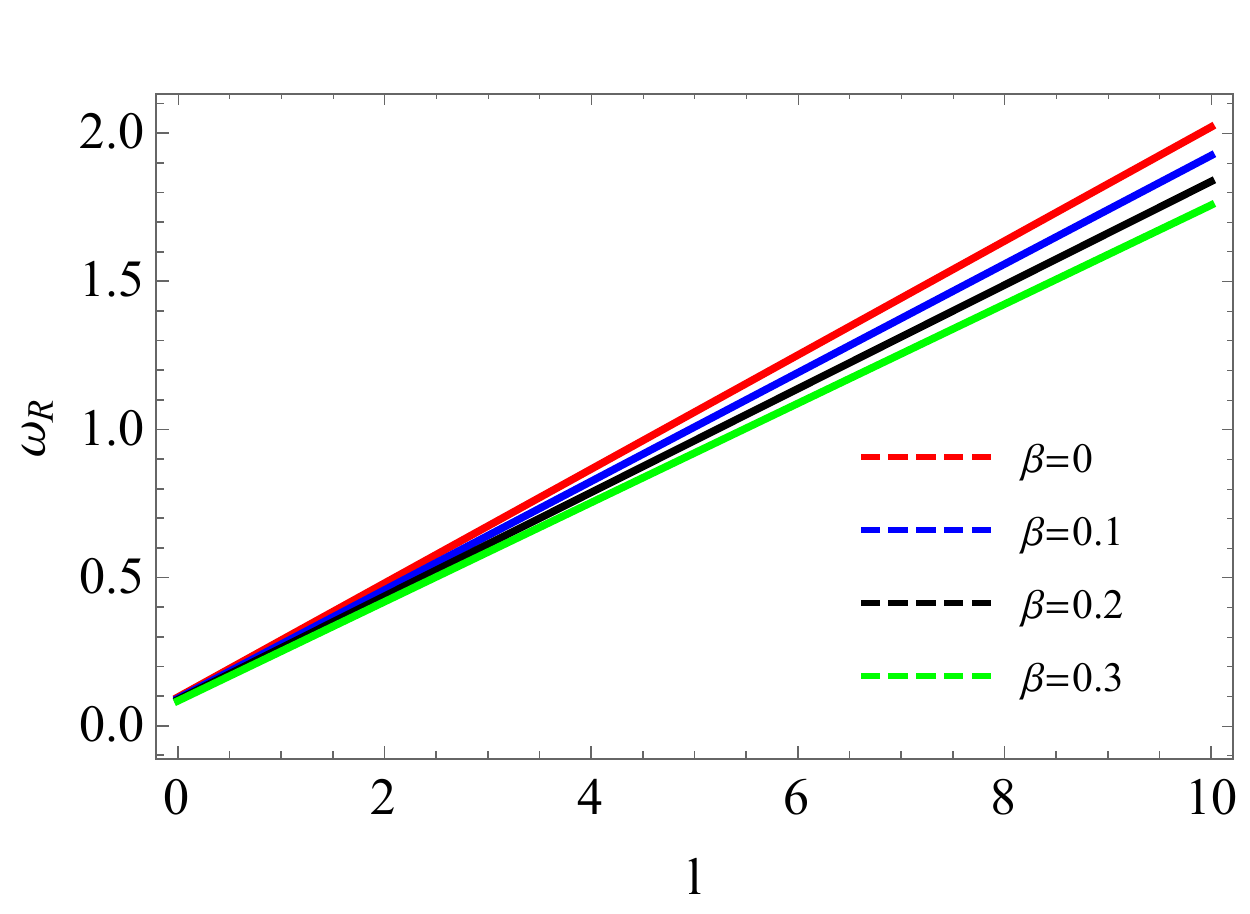}
\caption{Left panel: The real part of QNMs as a function of $\beta$ for a given $l$. Right panel: The real part of QNMs as a function of $l$ for a fixed $\beta$.  } \label{B}
\end{figure*}
The importance of this correspondence relies on the fact that the 
shadow radius represents an observable quantity which can be measured 
by using direct astronomical measurement. Therefore, it is more 
convenient to express the real part of the QNMs frequencies in terms 
of the black hole shadow radius instead of the angular velocity. Another 
advantage of using \eqref{k1} is the possibility to determine the shadow 
radius once we have calculated the real part of QNMs and this, in turn, 
does not necessitate the use of the standard geodesic method. This 
close connection could be understood from the fact that the 
gravitational waves can be treated as massless particles propagating 
along the last null unstable orbit and out to infinity. It is thus 
expected that, in the eikonal limit, this correspondence could be valid 
for the scalar, the electromagnetic, and the gravitational field 
perturbations because they have the same behavior. 

Although the relation~\eqref{k1} is accurate only for large $l$, this 
relation can provide valuable information regarding the effect of the 
electric charge $q_e$ on the shadow radius even for small $l$.To illustrate this fact,  we can use the correspondence between the shadow radius of the black hole and the real part of QNMs to sub-leading regime to half of its value  reported recently in Ref. \cite{Cuadros-Melgar:2020kqn}
\begin{equation}\label{modeq}
    \omega_{\Re}=R_s^{-1}\left(l+\frac{1}{2}\right).
\end{equation}
From the last equation it is clear that at high angular momentum regime, i.e. $l>>1$,  Eq. (\ref{k1}) is obtained. Again, this correspondence is accurate in the eikonal regime, but sometimes it is still accurate even for small multipole number $l$. The shadow radius for the static case reads

\begin{eqnarray}
    R_s=3 \sqrt{3} M\left(1+\frac{\beta}{2 M^2}\right)
\end{eqnarray}
In Table I we show the numerical values for the real part of QNMs obtained from the shadow radius.
In what follows we shall compare the above results with the results obtained via the WKB method.

\begin{widetext}
{\begin{center}
\begin{table}[tbp]
\begin{tabular}{|l|l|l|l|l|l|}
\hline
\multicolumn{1}{|c|}{ } &  \multicolumn{1}{c|}{  $l=1, n=0$ } & \multicolumn{1}{c|}{  $l=2, n=0$ } &   \multicolumn{1}{c|}{  $l=3, n=0$ } & \multicolumn{1}{c|}{  $l=4, n=0$ } & \multicolumn{1}{c|}{}\\\hline
  $\beta/M^2$ &\,\,\,\,$\omega_{\Re}$  &\,\,\,\,$\omega_{\Re}$ &\,\,\,\,$\omega_{\Re}$  &\,\,\,\,$\omega_{\Re}$   & \,\,\,\,$R_s [M]$\,\,\,\,   \\ \hline
0 & 0.288675 & 0.481125 & 0.673575  & 0.866025   &  $3 \sqrt{3}$ \\ 
0.1 & 0.274929  & 0.458214  & 0.641500 & 0.824786   & 5.45596 \\ 
0.2 & 0.262432 & 0.437387  & 0.612341 &  0.787296 & 5.71577 \\
0.3 & 0.251022 & 0.418370  & 0.585718   & 0.753066 & 5.97558  \\
0.4 & 0.240563 & 0.400938 &  0.561313 &  0.721688 &  6.23538 \\\hline
\end{tabular}
\caption{Numerical values for the shadow radius and the real part of QNMs obtained via Eq. (\ref{modeq}). }\label{table4}
\end{table}
\end{center}}
\end{widetext}

\subsection{Scalar field perturbations}
Let us consider the equation of motion for a massless scalar field which given by the Klein-Gordon equation and in the background of the curve spacetime can 
be written as follows
\begin{equation}
	\frac{1}{\sqrt{-g}} \partial_{\mu} \left(\sqrt{-g}\; g^{\mu \nu}\; 
	\partial_{\nu} \Phi \right) =0. \label{kgeq}
\end{equation}
Here $\Phi$ represents the massless scalar field and it is a function 
of coordinates $(t,r,\theta,\phi,\psi)$. We further consider an ansatz 
of the scalar field 
\begin{equation}
	\Phi(t,r,\theta,\phi,\psi) = \sum_{lm} e^{-i\omega t}\; 
	\frac{\Psi_l(r)}{r^{3/2}}\; Y_{lm}(r,\theta) , 
	\label{scfield}
\end{equation}
where $e^{-i\omega t} $ represents the time evolution of the field and 
$Y_{lm}(r,\theta)$ denotes the spherical harmonics function. Plunging 
the ansatz \eqref{scfield} into \eqref{kgeq} and applying the separation 
of variables method, we obtain the standard Schr{\"o}dinger-like wave 
equations
\begin{equation}
	\frac{d^2\Psi_l (r_{\ast})}{dr_{\ast}^2} 
	+\left(\omega^2 -V_s(r_{\ast})\right) \Psi_l(r_{\ast})=0, 
	\label{weq}
\end{equation}
where $\omega$ is the frequency of the perturbation and $r_{\ast}$ 
represents the tortoise coordinates having the relation
\begin{equation}
	dr_{\ast}=\frac{dr}{f(r)} \Rightarrow r_{\ast} = \int \frac{dr}{f(r)}.
\end{equation}
The advantage of using the tortoise coordinates here is to extend the 
range used in a survey of the QNMs. The tortoise coordinate is being 
mapped the semi-infinite region from the horizon to infinity into 
($-\infty,+\infty$) region. The effective potential in gives
\begin{eqnarray}\notag
	V_s(r_{\ast}) &=& \left[1-\frac{2M}{r}\left(1+\frac{\beta}{2 M^2}\right)\right]\\
	& \times & \left[\frac{l(l+1)}{r^2} + \frac{2 M}{r^3}\left(1+\frac{\beta}{2 M^2}\right)\right] 
\end{eqnarray}
where $l$ denotes the multipole number. On having the expression of the effective potential in our hand, 
we are now in a position to apply the WKB approach in order to compute 
the QNMs due to the scalar field perturbations. In our study, we are going 
to consider the sixth-order WKB method which is developed by Konoplya. 
 On the other hand, we find that an 
increase of $\beta$ the real part of QNMs decreases (cf. Table II). This indicates that the scalar field 
perturbations in a GUP modified black hole oscillate more
slowly compared to the Schwarzschild vacuum
black holes. 

\subsection{Electromagnetic field perturbations}
In the case of  the electromagnetic field perturbations we have the field equation
\begin{equation}
	\frac{1}{\sqrt{-g}}\partial_{\mu}\left[ \sqrt{-g}\;	g^{\lambda \mu}\; g^{\sigma \nu} 
	\left( \partial_{\lambda} A_{\sigma} -\partial_{\sigma} A_{\lambda} 
	\right)	\right]=0. \label{maxeqs}
\end{equation}
Without going into details, after we substitute all related expressions into \eqref{maxeqs} and using the 
standard tortoise coordinate transformation $dr_{\ast} =dr/f(r)$ yielding the effective potential
\begin{equation}
V_E(r)=\left[1-\frac{2M}{r}\left(1+\frac{\beta}{2 M^2}\right)\right] \frac{l(l+1)}{r^2}.
\end{equation}

In Table III we show the numerical results obtained via WKB method. Similarly to the previous case, the real part of QNMs decreases.

\subsection{Gravitational field perturbations}
Our final example will be the study of gravitational field perturbations. Before writing the field equation let us recall that the general form of the perturbed metric is given by
\begin{eqnarray}\notag
ds^2&=&-e^{2\nu}dt^2+e^{2\psi}(d\phi-\sigma dt-q_rdr-q_\theta d\theta)^2\\
&+& e^{-2\mu_2}dr^2+e^{-2\mu_3}d\theta^2,
\end{eqnarray}
in which $e^{2\nu}=e^{2\mu_2}=f(r)$, $e^{2\mu_3}=r^2, e^{2\psi}=r^2\sin^2\theta$ and $\sigma=q_r=q_\theta=0$ for non-perturbed case. The perturbations will lead to non-vanishing values of $\sigma, q_r, q_\theta$ and increments in $\nu, \mu_2, \mu_3, \psi$, which are corresponding to axial and polar perturbations, respectively. Here we shall consider the axial type ones. The perturbation equation reads
\begin{equation}
r^4\frac{\partial}{\partial r}\Big(\frac{f(r)}{r^2}\frac{\partial Q}{\partial r}\Big)+\sin^3\theta \frac{\partial}{\partial \theta}\Big(\frac{1}{\sin^3\theta}\frac{\partial Q}{\partial \theta}\Big)-\frac{r^2}{f(r)}\frac{\partial^2 Q}{\partial t^2}=0,
\end{equation}
where
\begin{eqnarray}\notag
Q(t,r,\theta)&=& e^{i\omega t}Q(r,\theta),\\\notag
Q(r,\theta)&=&r^2f(r)\sin^3\theta Q_{r\theta},\\
 Q_{r\theta}&=& q_{r,\theta}-q_{\theta,r}.
\end{eqnarray}
Further with $Q(r,\theta)=r\Psi(r)C^{-2/3}_{l+2}$, it can be reduced to Schrodinger wave-like equations:
\begin{eqnarray}\label{schrodinger}
\frac{d^2\Psi}{dr_*^2}+[\omega^2-V_G(r)]\Psi=0,\;\;dr_*=f(r)dr,
\end{eqnarray}
for gravitational field $\Psi$.
The effective potentials take the form as:
\begin{eqnarray}\notag
V_G(r)&=&\left[1-\frac{2M}{r}\left(1+\frac{\beta}{2 M^2}\right)\right]\Big[\frac{l(l+1)}{r^2}\\
&-& \frac{6 M}{r^3}\left(1+\frac{\beta}{2 M^2}\right) \Big].
\end{eqnarray}
In Table IV, we show the effect of GUP parameter on QNMs frequencies.  Note that the oscillation frequency $f$ depend on the black
hole parameters by converting the frequencies calculated in geometrical units into kHz, one should multiply $\omega$ by $2 \pi\, (5.142 \text{kHz})M_{\odot}/M$. For example, the first gravitational quasinormal mode frequency of
a Schwarzschild black hole corresponds to the fundamental $n = 0$ quadrupole $l=2$ mode and it is
$ \omega\,M \simeq 0.3736 - 0.089 \,i$, where we measure in units if the black hole mass $M=1$. 
For a black hole of 10 solar masses for $\beta/M^2=\{0,0.1,0.2,0.3,0.4\}$ we have the oscillation frequency is $f=\{1.2,1.16,1.12,1.08,1.04 \}$ kHz. This shows that the oscillation frequency due to the GUP effect indeed decreases.

\begin{widetext}
{\begin{center}
\begin{table}[h]
	\begin{tabular}{|l|l|l|l|l|}
		\hline
	\multicolumn{1}{|c|}{  } &  \multicolumn{1}{c|}{  $l=1, n=0$ } & 
	\multicolumn{1}{c|}{ $l=2, n=0$ } & \multicolumn{1}{c|}{ $l=3, n=0$} & \multicolumn{1}{c|}{ $l=4, n=0$}\\
		\hline
	$\beta/M^2$ & \,\,\,\,\,\,\,$\omega \,(WKB)$ & \,\,\,\,\,\,\,$\omega \,(WKB)$ & \,\,\,\,\,\,\,$\omega \,(WKB)$  & \,\,\,\,\,\,\,$\omega \,(WKB)$  \\ 
	\hline
	0 	& 0.29291 - 0.0977616 i & 0.483642 - 0.0967661  i & 0.675366 - 0.0965006  i & 0.867416 - 0.0963919   i  \\
	0.1 & 0.278962 - 0.0931063  i & 0.460611 - 0.0921582  i & 0.643206 - 0.0919053  i & 0.82611 - 0.0918018 i \\
	0.2 & 0.266282 - 0.0888742 i & 0.439674 - 0.0879692 i & 0.613969 - 0.0877278 i  & 0.78856 - 0.087629  i\\
	0.3 & 0.254704 - 0.0850101 i & 0.420558 - 0.0841444 i & 0.587275 - 0.0839136  i  & 0.754274 - 0.083819 i \\
   0.4 & 0.244091 - 0.081468  i & 0.403035 - 0.0806384 i & 0.562805 - 0.0804172   i & 0.722846 - 0.0803266 i\\ 
    \hline
	\end{tabular}
		\caption{\label{table1} Real and imaginary parts of the QNMs frequencies 
	in scalar field perturbations evaluated by WKB method up to the sixth order ($M=1$).}
\end{table}
\end{center}}
{\begin{center}
\begin{table}[h]
	\begin{tabular}{|l|l|l|l|l|}
		\hline
	\multicolumn{1}{|c|}{  } &  \multicolumn{1}{c|}{  $l=1, n=0$ } & 
	\multicolumn{1}{c|}{ $l=2, n=0$ } & \multicolumn{1}{c|}{ $l=3, n=0$} & \multicolumn{1}{c|}{ $l=4, n=0$}\\
		\hline
	$\beta/M^2$ & \,\,\,\,\,\,\,$\omega \,(WKB)$ & \,\,\,\,\,\,\,$\omega \,(WKB)$ & \,\,\,\,\,\,\,$\omega \,(WKB)$  & \,\,\,\,\,\,\,$\omega \,(WKB)$  \\ 
	\hline
	0 	& 0.248191 - 0.092637 i & 0.457593 - 0.0950112  i & 0.656898 - 0.0956171 i & 0.853095 - 0.0958601 i  \\
	0.1 & 0.236373 - 0.0882257 i &0.435803 - 0.0904868  i & 0.625618 - 0.0910639 i & 0.812472 - 0.0912954 i \\
	0.2 & 0.225629 - 0.0842155 i & 0.415994 - 0.0863738 i & 0.59718 - 0.0869247 i  & 0.775541 - 0.0871456  i\\
	0.3 & 0.215819 - 0.080554 i &0.397907 - 0.0826184 i & 0.571216 - 0.0831453 i  & 0.741822 - 0.0833566 i \\
    0.4 & 0.206826 - 0.0771975  i & 0.381328 - 0.079176  i & 0.547415 - 0.0796809  i &  0.710913 - 0.0798834 i\\ 
    \hline
	\end{tabular}
		\caption{\label{table1} Real and imaginary parts of the QNMs frequencies 
	in electromagnetic field perturbations  evaluated by WKB method up to the sixth order ($M=1$).}
\end{table}
\end{center}}
{\begin{center}
\begin{table}[h]
	\begin{tabular}{|l|l|l|l|l|}
		\hline
	\multicolumn{1}{|c|}{  } &  \multicolumn{1}{c|}{  $l=2, n=0$ } & 
	\multicolumn{1}{c|}{ $l=3, n=0$ } & \multicolumn{1}{c|}{ $l=4, n=0$} & \multicolumn{1}{c|}{ $l=5, n=0$}\\
		\hline
	$\beta/M^2$ & \,\,\,\,\,\,\,$\omega \,(WKB)$ & \,\,\,\,\,\,\,$\omega \,(WKB)$ & \,\,\,\,\,\,\,$\omega \,(WKB)$  & \,\,\,\,\,\,\,$\omega \,(WKB)$  \\ 
	\hline
	0	& 0.373619 - 0.088891 i  & 0.599443 - 0.0927025 i & 0.809178 - 0.0941641 i& 1.0123 - 0.0948706 i \\
	0.1 	& 0.359825 - 0.0849172 i & 0.573568 - 0.0884261 i& 0.772668 - 0.0897598 i & 0.965723 - 0.0904047 i \\
	0.2 	& 0.34692 - 0.0812921 i  & 0.549809 - 0.0845263 i & 0.739299 - 0.0857488 i& 0.923241 - 0.0863402 i \\
	0.3 	& 0.334836 - 0.0779684  i & 0.527919 - 0.0809554 i & 0.708684 - 0.0820807 i& 0.884335 - 0.0826253 i \\
	0.4 	& 0.32351 - 0.0749082 i  & 0.507689 - 0.0776736 i & 0.680497 - 0.0787133 i & 0.848571 - 0.0792168 i  \\
    \hline
	\end{tabular}
		\caption{\label{table1} Real and imaginary parts of the QNMs frequencies 
	in gravitational field perturbations  evaluated by WKB method up to the sixth order ($M=1$).}
\end{table}
\end{center}}
\end{widetext}

If we compare now the results obtained by means of the shadow radius and the WKB method we see that the agreement is very good even in the limit of small $l$. In fact we get better agreement for scalar field perturbations. In the limit $l>>1$, the agreement become exact, i.e. eikonal limit. This is very interesting result, having an experimental result for the shadow radius of a given black hole allows us to estimate the frequency of the gravitational waves if the black hole is perturbed. However, sometimes it may be useful to do the opposite; namely having an experimental result for a detected gravity wave, allows us to estimate the shadow radius of that black hole.

\section{Quasi-periodic oscillations}
Introducing the relevant universal constants, the metric~\eqref{metric} takes the form
\begin{align}
&g_{t t}= -c^2 \Big(1-\frac{2 G M r}{c^2 \zeta  (r^2+a^2 \zeta ^2 \cos ^2\theta )}\Big), \nonumber\\
&g_{t \phi }=-\frac{2 a G M r \sin ^2\theta }{c (r^2+a^2
	\zeta ^2 \cos ^2\theta )},\;g_{\theta  \theta }=r^2+a^2 \zeta ^2 \cos ^2\theta ,\nonumber\\
&g_{r r}= \frac{c^2 \zeta  (r^2+a^2 \zeta ^2 \cos ^2\theta )}{c^2 \zeta  (r^2+a^2 \zeta ^2)-2 G M r},\nonumber\\
\label{metric2}&g_{\phi  \phi }=\Big(r^2+a^2 \zeta ^2+\frac{2 a^2 G M r \zeta  \sin ^2\theta }{c^2 (r^2+a^2 \zeta ^2 \cos ^2\theta )}\Big) \sin ^2\theta ,
\end{align}
where $\zeta$ is expressed in terms of the dimensionless parameter $\beta_0$ as
\begin{equation}\label{zb0}
\zeta =\frac{2}{2+\beta_0}\qquad\text{and}\qquad \beta_0\equiv \frac{\beta}{M^2}.
\end{equation}
Recall that for the Kerr BH $\zeta=1$ and $\beta_0=0$ and that $\zeta(\beta_0)$ is a decreasing function of $\beta_0$.

For the numerical calculations to be carried out in this section, we take $M_\odot=1.9888\times 10^{30}$ (solar mass), $G=6.673\times 10^{-11}$ (gravitational constant), and $c=299792458$ (speed of light in vacuum) all given in SI units. These same constants will be written explicitly in some subsequent formulas of this section.

\begin{figure}[h]
	\centering
	\includegraphics[width=0.49\textwidth]{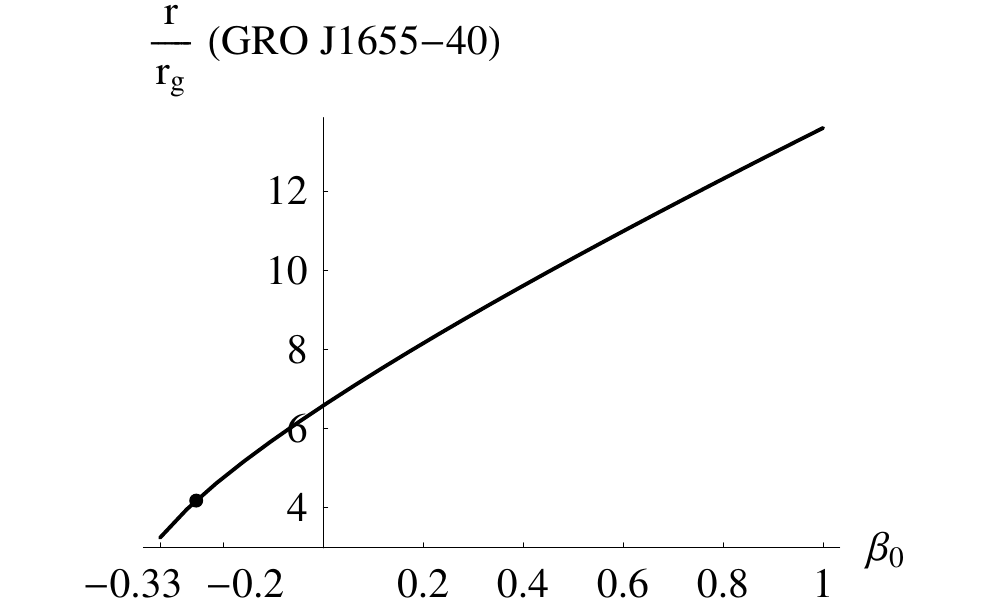}
	\caption{Plot of the dimensionless radius of the circle $u=r/r_g$~\eqref{uzeta}, where the 3/2 resonance occurs nearest the isco, versus the dimensionless parameter $\beta_0$~\eqref{zb0} for the the microquasar GRO J1655-40 taking $a_0=0.70$ [which is the intermediate value~\eqref{pr1}]. There is another increasing $u(\beta_0)$ branch but concave up (not shown in this plot) which provides higher values for $u$. Such a branch is not favorable since we believe that the resonance occurs nearest the isco. The point $(\beta_0,\,u)=(-0.254251,\,4.1748)$, where the lower $\nu_L=300$ Hz QPO and the upper $\nu_U=450$ Hz QPO occur, is shown by the black spot.}\label{FigGROub}
\end{figure}
\begin{figure}[h]
	\centering
	\includegraphics[width=0.49\textwidth]{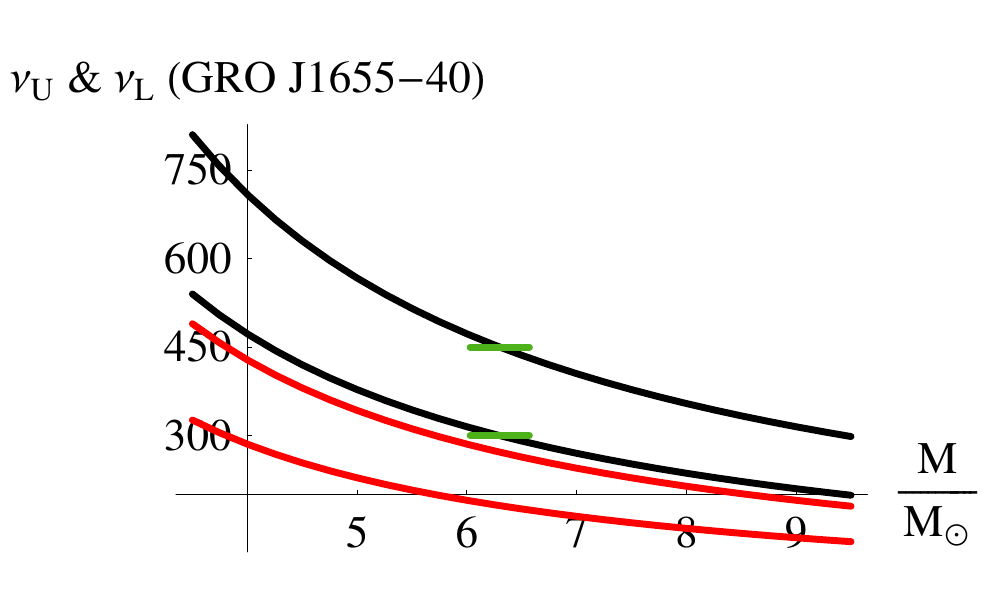}
	\caption{Fitting the particle oscillation upper and lower frequencies to the observed frequencies (in Hz) for the microquasar GRO J1655-40 at the 3/2 resonance radius. In the black plots the microquasar is treated as a GUP-modified Kerr BH given by~\eqref{metric2} with $\beta_0=-0.254251$ ($\zeta=1.14564$). The upper black curve represents $\nu_U=\nu_\theta$ and the lower black curve represents $\nu_L=\nu_r$ with $\nu_U/\nu_L=3/2$, and the green curve represent the mass error band as given in~\eqref{pr1}. The black curves cross the mass error bands ensuring a good curve fitting. In the red plots the microquasar is treated as a Kerr BH $\beta_0=0$ ($\zeta=1$). Since the red plots do not cross the mass error bands, a description of the astrophysical object by a Kerr BH fails to justify the occurrence of the 3/2 resonance.}\label{FigGRO}
\end{figure}
\begin{figure}[h]
	\centering
	\includegraphics[width=0.49\textwidth]{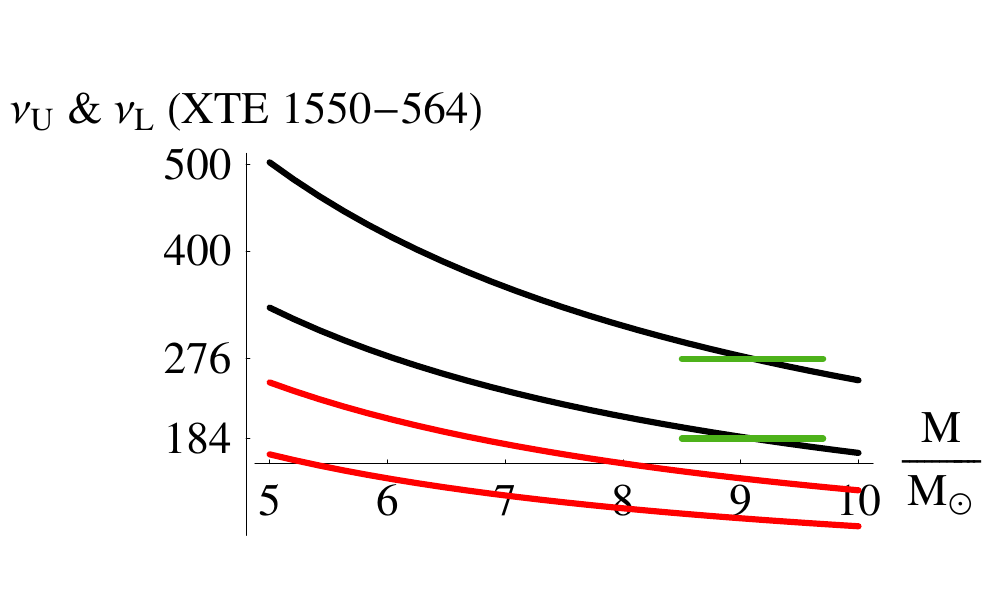}
	\caption{Fitting the particle oscillation upper and lower frequencies to the observed frequencies (in Hz) for the microquasar XTE J1550-564 at the 3/2 resonance radius. In the black plots the microquasar is treated as a GUP-modified Kerr BH given by~\eqref{metric2} with $\beta_0=-0.532673$ ($\zeta=1.36302$). The upper black curve represents $\nu_U=\nu_\theta$ and the lower black curve represents $\nu_L=\nu_r$ with $\nu_U/\nu_L=3/2$, and the green curve represent the mass error band as given in~\eqref{pr2}. The black curves cross the mass error bands ensuring a good curve fitting. In the red plots the microquasar is treated as a Kerr BH $\beta_0=0$ ($\zeta=1$). Since the red plots do not cross the mass error bands, a description of the astrophysical object by a Kerr BH fails to justify the occurrence of the 3/2 resonance.}\label{FigXTE}
\end{figure}
\begin{figure}[h]
	\centering
	\includegraphics[width=0.49\textwidth]{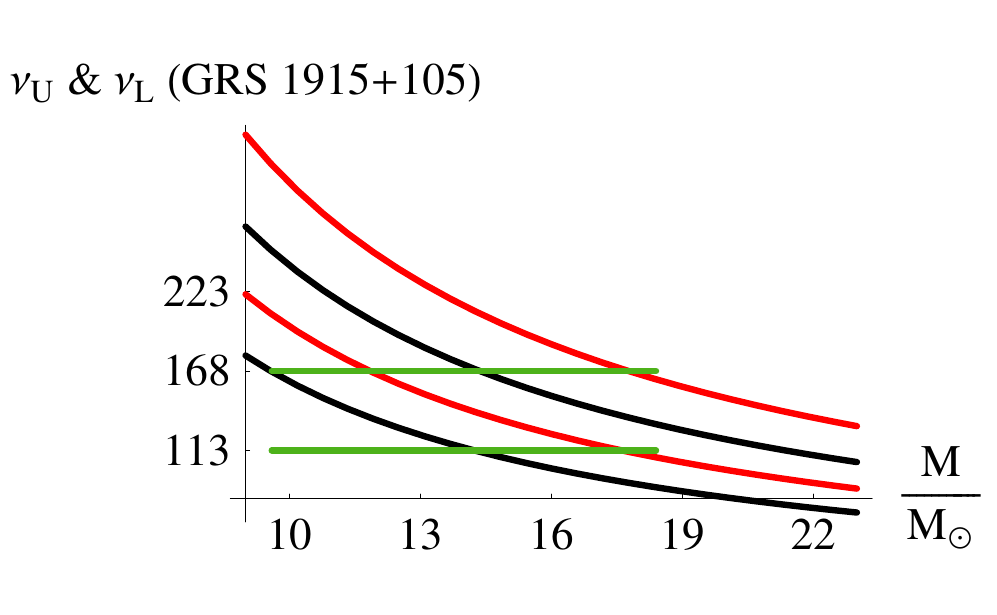}
	\caption{Fitting the particle oscillation upper and lower frequencies to the observed frequencies (in Hz) for the microquasar GRS 1915+105 at the 3/2 resonance radius. In the black plots the microquasar is treated as a GUP-modified Kerr BH given by~\eqref{metric2} with $\beta_0=0.071347$ ($\zeta=0.965555$). The upper black curve represents $\nu_U=\nu_\theta$ and the lower black curve represents $\nu_L=\nu_r$ with $\nu_U/\nu_L=3/2$, and the green curve represent the mass error band as given in~\eqref{pr2}. The black curves cross the mass error bands well in the middle ensuring a good curve fitting. In the red plots the microquasar is treated as a Kerr BH $\beta_0=0$ ($\zeta=1$). The red plots also cross the mass error bands but at the rightmost points. A description of the astrophysical object by a Kerr BH does not provide a good curve fitting to justify the occurrence of the 3/2 resonance.}\label{FigGRS}
\end{figure}

In the power spectra of Fig.~3 of Ref.~\cite{res}, we clearly see two peaks at 300 Hz and 450 Hz, representing, respectively, the possible occurrence of the lower $\nu_L=300$ Hz quasi-periodic oscillation (QPO), and of the upper $\nu_U=450$ Hz QPO from the Galactic microquasar GRO J1655-40. Similar peaks have been obtained for the microquasars XTE J1550-564 and GRS 1915+105 obeying the remarkable relation, $\nu_U/\nu_L=3/2$~\cite{qpos1}. Some of the physical {quantities} of these three microquasars and their uncertainties are as follows~\cite{res,res2}:
\begin{multline}\label{pr1}
\text{GRO J1655-40 : }\frac{M}{M_\odot}=6.30\pm 0.27,\;\frac{a}{r_g}=0.70\pm 0.05\\
\nu_U=450\pm 3 \text{ Hz},\;\nu_L=300\pm 5 \text{ Hz},
\end{multline}
\begin{multline}\label{pr2}
\text{XTE J1550-564 : }\frac{M}{M_\odot}=9.1\pm 0.6,\;\frac{a}{r_g}=0.405\pm 0.115\\
\nu_U=276\pm 3 \text{ Hz},\;\nu_L=184\pm 5 \text{ Hz},
\end{multline}
\begin{multline}\label{pr3}
\text{GRS 1915+105 : }\frac{M}{M_\odot}=14.0\pm 4.4,\;\frac{a}{r_g}=0.99\pm 0.01\\
\nu_U=168\pm 3 \text{ Hz},\;\nu_L=113\pm 5 \text{ Hz},
\end{multline}
where $r_g\equiv GM/c^2$.

These twin values of the QPOs are most certainly due to the phenomenon of resonance which occurs in the vicinity of the ISCO, where the in-falling charged particles perform radial and vertical oscillations around almost circular orbits. The local radial and vertical oscillations are denoted by ($\Omega_r,\,\Omega_\theta$), respectively. These two oscillations couple generally non-linearly to yield resonances in the power spectra~\cite{res3,res4}. For the case of uncharged rotating BH ($\Omega_r,\,\Omega_\theta$) are given by~\cite{qposknb} (see also~\cite{see1,see2})
\begin{align}\label{Omega}
&\Omega_{r}^2\equiv (\partial_{r}\Gamma^{r}_{ij}-4\Gamma^{r}_{ik}\Gamma^{k}_{rj})u^iu^j,\qquad (i,\,j,\,k=t,\,\phi),\nonumber\\
&\Omega_{\theta}^2\equiv (\partial_{\theta}\Gamma^{\theta}_{ij})u^iu^j,\qquad (i,\,j=t,\,\phi),
\end{align}
In obtaining these expressions we assumed that the main motion of the particle is circular in the equatorial plane ($\theta=\pi/2$) where the particle exhibits radial and vertical oscillations. The circular motion is stable only if $\Omega_r^2>0$ and $\Omega_\theta^2>0$. In the equatorial plane the four-velocity vector of the particle has only two nonzero components $u^{\mu}=(u^t,\,0,\,0,\,u^{\phi})=u^t(1,\,0,\,0,\,\omega)$, where $\omega=d \phi/d t$ is the angular velocity of the test particle. They are related by~\cite{qposknb}
\begin{align}\label{velocity}
&\omega =\frac{-\partial _r g_{t \phi }\pm \sqrt{\left(\partial _r g_{t \phi }\right)^2-\partial _r g_{t t} \partial _r g_{\phi  \phi }}}{\partial _r g_{\phi
		\phi }},\nonumber\\
&u^t=\frac{c}{\sqrt{-\left(g_{t t}+2 \partial _r g_{t \phi } \omega +g_{\phi  \phi } \omega ^2\right)}},\nonumber\\
&u^{\phi }=\omega  u^t .
\end{align}
In these expressions the summations extend over ($t,\,\phi$). It is understood that all the functions appearing in~\eqref{Omega} and~\eqref{velocity} are evaluated at $\theta=\pi/2$.

The locally measured frequencies ($\Omega_{r},\,\Omega_{\theta}$) are related to the spatially-remote observer's frequencies ($\nu_{r},\,\nu_{\theta}$) by
\begin{align}\label{pr4}
&\nu_r=\frac{1}{2\pi}~\frac{1}{u^t}~\Omega_{r},
&\nu_\theta=\frac{1}{2\pi}~\frac{1}{u^t}~\Omega_{\theta}.
\end{align}
Using the form~\eqref{metric2} of the metric and introducing the dimensionless parameters $y$ and $a_0$ defined by
\begin{equation}\label{rg}
u\equiv\frac{r}{r_g},\quad a_0\equiv\frac{a}{r_g},\quad r_g\equiv\frac{GM}{c^2},
\end{equation}
we arrive at
\begin{align}\label{nrt}
&\nu _r=\frac{c^3(u^{3/2}-a_0 \sqrt{\zeta }) \sqrt{\zeta  u^2-6 u+8 a_0 \zeta ^{3/2} \sqrt{u}-3 a_0^2 \zeta ^3}}{2 \pi  G M\zeta  (u^3-a_0^2\zeta ) u},\\
&\nu _{\theta }=\frac{c^3(u^{3/2}-a_0 \sqrt{\zeta }) \sqrt{\zeta u^2-4 a_0 \zeta ^{3/2} \sqrt{u}+3 a_0^2 \zeta ^3}}{2 \pi  G M\zeta (u^3-a_0^2 \zeta ) u}.\nonumber
\end{align}
Setting $\zeta=1$ these expressions reduce to the corresponding expressions for the Kerr BH. 

On confronting the observed ratio $\nu_U/\nu_L=3/2$ most workers in this field appeal to parametric resonance to explain the observed ratio assuming that $\nu_{\theta }/\nu_{r}=n/2$ and $n\in \mathbb{N}^+$. In almost all applications of parametric resonance one considers the case $n=1$~\cite{b1,b2,b3,b4} where in this case $\nu _{r}$ is the natural frequency of the system and $\nu_{\theta }$ is the parametric excitation ($T_{\theta }=2T_{r}$, the corresponding periods), that is, the vertical oscillations supply energy to the radial oscillations causing resonance~\cite{b4}. However, since always $\nu_{\theta }>\nu_{r}$, it is neither possible to have $n=1$ nor $n=2$ in the vicinity of the isco where it is thought that the resonance effects take place. The next allowed choice is thus $n=3$ by which $\nu_{r}$ becomes the parametric excitation that supplies energy to the vertical oscillations. In this work we work with the ansatz $\nu_U=\nu_\theta$, $\nu_L=\nu_r$ along with $\nu_{\theta }/\nu_{r}=3/2$ ($n=3$). Equation~\eqref{nrt} yields
\begin{equation}\label{ratio}
\Big(\frac{\nu_{U}}{\nu_{L}}\Big)^2 =\frac{\zeta u^2-4 a_0 \zeta ^{3/2} \sqrt{u}+3 a_0^2 \zeta ^3}{\zeta  u^2-6 u+8 a_0 \zeta ^{3/2} \sqrt{u}-3 a_0^2 \zeta ^3}=\frac{9}{4}.
\end{equation}
Expressing $u$, where the 3/2 resonance occurs nearest the isco, in terms of ($a_0,\,\zeta$) we obtain
\begin{align}\label{uzeta}
&u=\frac{1}{5 \zeta } \bigg[27+Z-\sqrt{486+130 a_0^2 \zeta ^4-\frac{Y}{X}-X+\frac{9680 a_0^2 \zeta ^4}{Z}}\bigg],\nonumber\\
&X=\Big[14348907-5a_0^2 \zeta ^4[4048137-1259725 a_0^2 \zeta ^4-54925 a_0^4 \zeta ^8] \nonumber\\
&\qquad +24200 a_0^3 \zeta ^{6}\sqrt{12393-23378 a_0^2
	\zeta ^4+10985 a_0^4 \zeta ^8}\Big]^{1/3},\nonumber\\
&Y=59049-55530 a_0^2 \zeta ^4+4225 a_0^4 \zeta ^8,\nonumber\\
&Z=\sqrt{243+65 a_0^2 \zeta ^4+\frac{Y}{X}+X}.
\end{align}

For fixed $a_0$ this provides a relation between the dimensionless parameter $\beta_0$~\eqref{zb0} and the dimensionless radius of the circle $u$ where the 3/2 resonance occurs nearest the isco. For the microquasar GRO J1655-40 we take $a_0=0.70$, which is the intermediate value~\eqref{pr1}, and we plot $u=r/r_g$ in terms of $\beta_0$ as shown in Fig.~\ref{FigGROub}, which an increasing function of $\beta_0$ and concave down. There is another increasing $u(\beta_0)$ branch but concave up (not shown in Fig.~\ref{FigGROub}) which provides higher values for $u$. Such a branch is not favorable since we believe that the resonance occurs nearest the isco.  For the other two microquasars~\eqref{pr2} and~\eqref{pr3} we have obtained similar increasing $u(\beta_0)$ functions. Keeping only one branch of $u(\beta_0)$, now to determine the unique values of ($\beta_0,\,u$) corresponding to the occurrence of the upper $\nu_{U}$ and lower $\nu_{L}$ we solve the equation  $\nu_{U}=\text{observed value}$ (or the equation $\nu_{L}=\text{observed value}$) for each microquasar, where the observed values of ($\nu_{U},\,\nu_{L}$) for the three microquasars are given in~\eqref{pr1}, \eqref{pr2} and~\eqref{pr3}. For the three microquasars taking ($M,\,a_0$) to be the intermediate value given in~\eqref{pr1}, \eqref{pr2} and~\eqref{pr3}, we obtained
\begin{align}\label{coord}
&\text{GRO J1655-40 :}& &\beta_0=-0.254251,& & u=4.1748 ,\nonumber\\
&\text{XTE J1550-564 :}& &\beta_0=-0.532673,& & u=4.54216 ,\\
&\text{GRS 1915+105 :}& &\beta_0=0.071347,& & u=5.0168 .\nonumber
\end{align}
Recall that isco is defined by $\Omega_{r}(u_\text{isco})=0$ and these values of $u$ are certainly higher than $u_\text{isco}$~\cite{qposknb} (see also~\cite{see1,see2}). In Fig.~\ref{FigGROub}, corresponding to the microquasar GRO J1655-40, the point $(\beta_0,\,u)=(-0.254251,\,4.1748)$, where the lower $\nu_L=300$ Hz QPO and the upper $\nu_U=450$ Hz QPO occur, is shown by the black spot.

Perfect curve fitting of the particle oscillation upper and lower frequencies to the observed frequencies (in Hz) for each microquasar are shown in Figs.~\ref{FigGRO}, \ref{FigXTE} and~\ref{FigGRS}. In the black plotss each microquasar is treated as a GUP-modified Kerr BH given by~\eqref{metric2} and taking the coordinates of $(\beta_0,\,u)$ to be the values given in~\eqref{coord}. Here $u$ is the dimensionless radius where the 3/2 resonance occurs. In the red plots the microquasar is treated as a Kerr BH $\beta_0=0$ ($\zeta=1$). We see from these three figures that the black plots cross the mass error bands exactly in the middle point. For the microquasars GRO J1655-40 and XTE J1550-564, the red plots do not cross the mass error bands, hence a description of the astrophysical object by a Kerr BH fails to justify the occurrence of the 3/2 resonance. For the microquasar GRS 1915+105, however, the red plots also cross the mass error bands but at the rightmost points. Knowing that for the microquasar GRS 1915+105 the mass error band is the largest, this provides a mediocre curve fitting of the particle oscillation upper and lower frequencies to the observed frequencies.

Based on the previous analysis, we restrict the values of $\beta_0$ to lie between the smallest and largest values we obtained above:
\begin{multline}\label{bz}
-0.532673\lesssim \beta_0\lesssim 0.071347\\
\Rightarrow 0.965555\lesssim \zeta \lesssim 1.36302 .
\end{multline}

Using the last result, we can determine the upper bound of GUP restoring the Planck mass via $\beta \lesssim 0.071347 \times M^2/M_P^2$. If we take the mass $M=14 \times M_\odot$, we find $\beta \lesssim 1.15 \times 10^{77}$. Although this is huge number, compared to the shadow case here we obtain a better constraint for $\beta$. 

\section{Conclusions}
In this paper we have studied the effect of the Generalised Uncertainty Principle (GUP) on the shadow of GUP-modified Kerr black hole and the correspondence between the shadow radius and the real part of the quasinormal modes  (QNMs). We have found  that the shadow curvature radius of the GUP-modfied Kerr black hole is bigger compared to the Kerr vacuum solution and increases linearly monotonically with the increase of the GUP parameter.Using the characteristic points of intrinsic curvature of the shadow er have calculated the angular size for these curvature radii of the shadow. Using the EHT data for the M87*black hole we have constraint the upper limit of the GUP parameter. Within $2\sigma $,  we have the upper and lower limit of the GUP parameter in terms of the interval $0 \lesssim \beta/M^2 \lesssim 0.78 $.

Finally, we have explored the connection between the shadow radius and the scalar/electromagnetic QNMs. It is argued that this correspondence works well even in the case of small $l$. This provides an interesting connection between the experimental data of the shadow and the detection of the gravity waves.  Having the shadow radius one can estimate the value of the QNMs frequency of a given black hole, or vice versa. 

We have shown that describing the microquasars GRO J1655-40, XTE J1550-564 and GRS 1915+105 as GUP-modified Kerr BHs yields perfect curve fitting of the particle oscillation upper and lower frequencies to the observed frequencies provided we restrict the values of the correction dimensionless parameter $\beta_0$ by $-0.532673\lesssim \beta_0\lesssim 0.071347$. These are very reasonable bounds knowing that the metric of the GUP-modified Kerr BH is a correction of the Kerr one where $\beta_0$ should lie in the vicinity of zero.

With the above main results, we would like to mention the above results can be extended to another type modified Kerr spacetime, the extended uncertainty principle (EUP) corrected black holes \cite{EUP}. In contrast to the GUP which modifies HUP at high energy regime, the EUP modifies HUP at low energy limits \cite{EUP}. In this way, it can be considered an effects at large scales.  Thus, it is interesting to explore the observational constraints on the EUP by using the observational data of shadow and QPOs. We will consider this issue in our future works.   

\subsection*{Acknowledgements}
Tao Zhu is supported by the National Key Research and Development Program of China Grant No.2020YFC2201503,  the Zhejiang Provincial Natural Science Foundation of China under Grants No. LR21A050001 and LY20A050002, the National Natural Science Foundation of China under Grant No. 11675143, and the Fundamental Research Funds for the Provincial Universities of Zhejiang in China under Grant No. RF-A2019015.


\begin{thebibliography}{99}

\bibitem{mage} M. Maggiore, Phys. Lett. B {\bf304}, 65 (1993).

\bibitem{vage} E. C.Vagenas, S. M. Alsaleh, A. F. Ali, Euro. Phys. Lett. {\bf120}, 40001 (2017).

\bibitem{vagee} A. F. Ali, S. Das and E.C. Vagenas, Phys. Lett. B {\bf678}, 497 (2009).

\bibitem{fras} A.M. Frassino and O. Panella, Phys. Rev. D {\bf85}, 045030 (2012).

\bibitem{j1} K. Jusufi, P. Channuie and M. Jamil, Eur. Phys. J. C {\bf80}, 127 (2020).

\bibitem{chen} P. Chen, R.J. Adler, Nucl. Phys. Proc. Suppl. {\bf124}, 103 (2003).

\bibitem{wu} W-Y. Wen and S-Y. Wu, Eur. Phys. J. C {\bf75}, 608 (2015).

\bibitem{spa} P. Nicolini, E. Spallucci and M. F. Wondrak, Phys. Lett. B {\bf797}, 134888 (2019).

\bibitem{min} E. Spallucci, A. Smailagic, {\it Advances in black holes research } p.1-26, Ed.: A. Barton, Nova Science Publisher, Inc. (2015), ISBN: 978-1-63463-168-6, arXiv:1410.1706 [gr-qc].

\bibitem{Carr:2020hiz} B. J. Carr, J. Mureika and P. Nicolini, JHEP {\bf 1507}, 052 (2015);  B.~Carr, H.~Mentzer, J.~Mureika, P.~Nicolini, Eur. Phys. J. C {\bf80},1166 (2020).

\bibitem{m87} J. C. S. Neves, Eur. Phys. J. C {\bf80}, 343 (2020).

\bibitem{const} \"O. \"Okc\"u and E. Aydiner, arXiv:2101.09524v1 [gr-qc]; F. Scardigli and R. Casadio, Eur. Phys. J. C {\bf75}, 425 (2015).

\bibitem{bambi} S. Nampalliwar and C. Bambi, {\it Tutorial Guide to X-ray and Gamma-ray Astronomy: Data Reduction and Analysis} (Ed. C. Bambi, Springer Singapore, 2020), arXiv:1810.07041.

\bibitem{bambi2} C. Bambi, {\it Black Holes: A Laboratory for Testing Strong Gravity} (Springer Singapore, Singapore, 2017).

\bibitem{eht} K. Akiyama et al. [Event Horizon Telescope Collaboration], Astrophys. J. 875, L1 (2019).

\bibitem{yuan} F. Yuan and R. Narayan, Annu. Rev. Astron. Astrophys. {\bf52}, 529 (2014).

\bibitem{jet}R. C. Walker, P. E. Hardee, F. B. Davies, C. Ly, and W. Junor,  Astrophys. J. {\bf855}, 128 (2018).

\bibitem{Wei:2018xks} S.~W.~Wei, Y.~X.~Liu and R.~B.~Mann, Phys. Rev. D \textbf{99}, 041303 (2019).

\bibitem{Allahyari:2019jqz} A.~Allahyari, M.~Khodadi, S.~Vagnozzi and D.~F.~Mota, JCAP \textbf{02}, 003 (2020).

\bibitem{cardoso} V. Cardoso, A. S. Miranda, E. Berti, H. Witek, and V. T. Zanchin, Phys. Rev. D {\bf79}, 064016 (2009).
 
\bibitem{Stefanov:2010xz} I.~Z.~Stefanov, S.~S.~Yazadjiev and G.~G.~Gyulchev,  Phys. Rev. Lett.  {\bf 104}, 251103 (2010).

\bibitem{Jusufi:2019ltj} K.~Jusufi, Phys. Rev. D  {\bf 101}, 084055 (2020).

\bibitem{Liu:2020ola} C.~Liu, T.~Zhu, Q.~Wu, K.~Jusufi, M.~Jamil, M.~Azreg-A\"{i}nou and A.~Wang,  Phys. Rev. D {\bf 101},  084001 (2020).

\bibitem{Cuadros-Melgar:2020kqn} B.~Cuadros-Melgar, R.~D.~B.~Fontana and J.~de Oliveira, [arXiv:2005.09761 [gr-qc]].


\bibitem{res} T. E.~Strohmayer, Astrophys. J. Lett. \textbf{552}, L49 (2001).

\bibitem{qpos1} J. E.~McClintock et al., Class. Quantum Grav. \textbf{28}, 114009 (2011).

\bibitem{res2} R. Shafee, J. E. McClintock, R. Narayan, S. W. Davis, L.-X. Li, and R. A. Remillard, Astrophys. J. Lett. \textbf{636}, L113 (2006).

\bibitem{res3} M. A. Abramowicz, V. Karas, W. Klu\'{z}niak, W. Lee, and P. Rebusco, Publ. Astron. Soc. Japan, \textbf{55}, 467 (2003).

\bibitem{res4} J. Hor\'{a}k and V. Karas, A{\&}A, \textbf{451}, 377 (2006).

\bibitem{qposknb} M.~Azreg-A\"{\i}nou, Int. J. Mod. Phys. D \textbf{28}, 1950013 (2019).

\bibitem{see1} K.~Jusufi, M.~Azreg-A\"{\i}nou, M.~Jamil, S.-W. Wei, Q.~Wu and A.~Wang, Phys. Rev. D {\bf103}, 024013 (2021)

\bibitem{see2} M. Ghasemi-Nodehi, M.~Azreg-A\"{\i}nou, K.~Jusufi and M.~Jamil, Phys. Rev. D {\bf102}, 104032 (2020).

\bibitem{b1} L.D. Landau and E.M. Lifshitz, \emph{Mechanics} (Pergamon Press, Oxford, 1976).

\bibitem{b2} A.H Nayfeh and D.T. Mook, \emph{Nonlinear Oscillations}, (Wiley-VCH Verlag GmbH, New Jersey, 1995).

\bibitem{b3} A. Lindner and D. Strauch, \emph{A Complete Course on Theoretical Physics: From Classical Mechanics to Advanced Quantum Statistics}, (Springer Nature Switzerland AG, 2018).

\bibitem{b4} E.I. Butikov, Parametric resonance, Computing in Science and Engineering (CiSE) May/June, 76 (1999).

\bibitem{EUP} J. R. Mureika, Phys. Lett. B {\bf789}, 88 (2019).



\end{thebibliography}
\end{document}